\documentclass[tighten]{aastex62}
\newcommand{\HI}{\text{H\hspace{0.15em}\textsc{i}}}
\newcommand{\HIs}{\text{\scriptsize H\hspace{0.105em}\textsc{i}}}
\newcommand{\NHI}{N_{\HIs}}
\newcommand{\NHIstar}{N_{\HIs}^{\ast}}
\newcommand{\WHI}{W_{\HIs}}

\newcommand{\tauHI}{\tau_{\HIs}}
\newcommand{\Ts}{T_{\text{s}}}
\newcommand{\XHI}{X_{\HIs}}
\newcommand{\THI}{T_{\HIs}}

\newcommand{\Htwo}{\text{H}_{2}}
\newcommand{\Htwos}{\text{{\scriptsize H}}_{2}}
\newcommand{\NHtwo}{N_{\Htwos}}

\newcommand{\HII}{\text{H\hspace{0.15em}\textsc{ii}}}
\newcommand{\Halpha}{\text{H}\alpha}

\newcommand{\NH}{N_{\text{H}}}

\newcommand{\NHref}{N_{\text{H,ref}}}

\newcommand{\COs}{\text{\scriptsize CO}}
\newcommand{\XCO}{X_{\COs}}

\newcommand{\WCO}{W_{\COs}}

\newcommand{\taud}{\tau_{\text{353}}}
\newcommand{\taudref}{\tau_{\text{353,ref}}}
\newcommand{\Td}{T_{\text{d}}}

\newcommand{\AJ}{A_{\it {\text{\it J}}}}

\newcommand{\Tbg}{T_{\text{bg}}}

\newcommand{\VLSR}{V_{\text{LSR}}}
\newcommand{\VHI}{V_{\text{\HI}}}

\newcommand{\Ucm}{\text{cm}}

\newcommand{\Ukm}{\text{km}}

\newcommand{\Us}{\text{s}}

\newcommand{\UK}{\text{K}}

\newcommand{\UCND}{\Ucm^{-2}} 

\newcommand{\UXCO}{\Ucm^{-2}\,\UK^{-1}\,\Ukm^{-1}\,\Us}

\newcommand{\Msolar}{M_{\sun}}

\accepted{May 8, 2019}

\begin{document}

\title{Gas and Dust Properties in the Chamaeleon Molecular Cloud Complex based on the Optically Thick $\HI$}


\author[0000-0001-6922-6583]{K. Hayashi}
\affiliation{Department of Physics, Nagoya University, Furo-cho, Chikusa-ku, Nagoya 464-8601, Japan}

\author{R. Okamoto}
\affiliation{Department of Physics, Nagoya University, Furo-cho, Chikusa-ku, Nagoya 464-8601, Japan}

\author{H. Yamamoto}
\affiliation{Department of Physics, Nagoya University, Furo-cho, Chikusa-ku, Nagoya 464-8601, Japan}

\author{T. Hayakawa}
\affiliation{Department of Physics, Nagoya University, Furo-cho, Chikusa-ku, Nagoya 464-8601, Japan}

\author[0000-0002-1411-5410]{K. Tachihara}
\affiliation{Department of Physics, Nagoya University, Furo-cho, Chikusa-ku, Nagoya 464-8601, Japan}

\author{Y. Fukui}
\affiliation{Institute for Advanced Research, Nagoya University, Furo-cho, Chikusa-ku, Nagoya 464-8601, Japan; sano@a.phys.nagoya-u.ac.jp}
\affiliation{Department of Physics, Nagoya University, Furo-cho, Chikusa-ku, Nagoya 464-8601, Japan}


\begin{abstract}

Gas and dust properties in the Chamaeleon molecular cloud complex have been investigated with emission lines from atomic hydrogen ($\HI$) and $^{12}$CO molecule, dust optical depth at 353 GHz ($\taud$), and $J$-band infrared extinction ($\AJ$).
We have found a scatter correlation between the $\HI$ integrated intensity ($\WHI$) and $\taud$ in the Chamaeleon region.
The scattering has been examined in terms of possible large optical depth in $\HI$ emission ($\tauHI$) using a total column density ($\NH$) model based on $\taud$.
A nonlinear relation of $\taud$ with the $\sim$1.2 power of $\AJ$ has been found in opaque regions ($\AJ \gtrsim 0.3$ mag), which may indicate dust evolution effect. 
If we apply this nonlinear relation to the $\NH$ model (i.e., $\NH \propto \taud^{1/1.2}$) allowing arbitrary $\tauHI$, the model curve reproduces well the $\WHI$--$\taud$ scatter correlation, suggesting optically thick $\HI$ ($\tauHI \sim$1.3) extended around the molecular clouds.
Based on the correlations between the CO integrated intensity and the $\NH$ model, we have then derived the CO-to-$\Htwo$ conversion factor ($\XCO$) on $\sim$1.5$^{\circ}$ scales (corresponding to $\sim$4 persec) and found spatial variations of $\XCO$ $\sim$(0.5--3)$\times$10$^{20}$ $\UXCO$ across the cloud complex, possibly depending on the radiation field inside or surrounding the molecular clouds.
These gas properties found in the Chamaeleon region are discussed through a comparison with other local molecular cloud complexes.

 \end{abstract}


\section{Introduction} \label{sec:Intro}

The neutral hydrogen on the atomic and molecular forms occupies major mass of the interstellar medium (ISM) and is a fundamental constituent of the ISM.
In the electronic ground state neutral atomic hydrogen ($\HI$) has two spin states where the spin angular momenta of a proton and an electron are parallel or anti-parallel.
The energy separation between the two states is small (5.9 $\times$ 10$^{-6}$ eV) and corresponds to a wavelength of 21 cm.
The 21 cm $\HI$ transition is used to calculate $\HI$ column density ($\NHI$) usually by assuming that the emission is completely optically thin (e.g., \citealt{BoulangerPerault88}).

The interstellar $\HI$ gas, however, has density and temperature which range over order of magnitude: density is distributed from 1~cm$^{-3}$ to 10$^{3}$~cm$^{-3}$ and temperature from 10~K to 10$^{4}$~K (e.g., \citealt{Draine11}). 
The gas consists of two distinct phases, the cold neutral medium (CNM) and the warm neutral medium (WNM). 
The CNM is dense and cool ($\sim$30~cm$^{-3}$ and $\sim$60~K), while the WNM is diffuse and warm ($\sim$0.6~cm$^{-3}$ and $\sim$2000~K) (e.g., \citealt{HeilesTroland03b}; \citealt{Draine11}).
The $\HI$ gas is highly turbulent and transient because it is continuously shocked by supernovae every million year.

To measure precisely the local $\HI$ gas with the large variations of the optical depth,
\cite{Fukui+14, Fukui+15} (hereafter F14, F15) proposed a method to calculate $\NHI$ based on dust optical depth at 353 GHz ($\taud$), which is estimated from modified black body spectra fitted to the fluxes at submillimeter wavelengths measured by the {\it Planck} and {\it IRAS} satellites \citep{Planck14a}.
$\taud$ is measured to be very small in the order of $\sim$10$^{-3}$, toward the Galactic mid plane and we are able to use $\taud$ as a tracer of $\NHI$ if the gas to dust ratio is uniform.
The results of F14 and F15 give a suggestion that the $\HI$ emission can be optically thick in the order of 1.0 with low spin temperature ($\Ts$ $\lesssim$ 100 K) and the considerable amount of atomic hydrogen is underestimated by the optically thin assumption often adopted in studies of the local ISM.

\citet{Stanimirovic+14} carried out $\HI$ absorption observations toward 26 radio continuum sources behind Perseus using the Arecibo 305 m telescope.
These authors showed that the $\HI$ optical depth ($\tauHI$) in the emission-absorption measurements is significantly smaller than that derived by F14 and F15; the peak optical depth of $\tauHI$~$>$~0.5 for only 21 out of 107 individual Gaussian  components, as opposed to F14 who found $\tauHI$~$>$~0.5 for 85\% of lines of sight at high Galactic latitudes.
The results by \citet{Stanimirovic+14} are consistent with those by \citet{HeilesTroland03a, HeilesTroland03b} toward 79 extragalactic sources and raised a question on F14 and F15.

Recently, \citet{Fukui+18} made synthetic observations of 21 cm emission and absorption by using the magnetohydrodynamic   simulations performed by \citet{InoueInutsuka12} and presented that the synthetic observations are consistent with the optically thick $\HI$ and the $\Ts$-dependent relationship between $\NHI$ and the $\HI$ integrated intensity ($\WHI$) suggested by F14 and F15. 
In addition, \citet{Fukui+18} found that the WNM with $\tauHI$ $<$ 0.5 and with $\Ts$ higher than 300~K is extended by $\sim$70\% in the sky and the radio absorption toward extragalactic continuum sources is biased toward WNM.
These results give an estimate that the optically thin approximation for the $\HI$ emission underestimates the $\HI$ mass by a factor of $\sim$1.3.
\cite{Okamoto+17} showed that the $\HI$ distribution in the Perseus molecular cloud can be reproduced well by the total column density ($\NH$) model as a function of $\sim$1/1.3th power of $\taud$. 
The authors derived that the $\HI$ column density is 1.6 times higher than that of the optically thin case, suggesting that a large amount of the optically thick $\HI$ around the molecular clouds.
The nonlinear behavior between $\taud$ and $\NH$ indicates dust evolution effect, as suggested from measurements of dust opacity in local molecular clouds (e.g., \cite{Roy+13} for Orion A molecular cloud).

Among other local molecular cloud complex, the Chamaeleon complex is known as a nearby low-mass star-forming region  at a distance of $\sim$150 pc, whose molecular cloud mass is estimated to be $\sim$5000--8300 $\Msolar$ (e.g., \citealt{Mizuno+01}; \citealt{Ackermann+12}; \citealt{Planck15}).
The cloud properties and the moderate star formation activities are reviewed by \cite{Luhman08}.
\citet{Planck15} attempted to model the gas distribution in the Chamaeleon region with a linear combination of $\HI$, CO and ``dark gas", a neutral gas component that cannot be traced by standard $\HI$ and CO observations, and estimated the gas column density by fitting these gas model maps to $\gamma$-rays and thermal dust emission models (dust extinction, $\taud$ and radiance).
The dark gas template is constructed through iterating fittings to the $\gamma$-rays and dust data alternately.
$\Ts$ in $\HI$ emission is assumed to be uniform and the optically thin approximation is adopted because it gives a better fit to the $\gamma$-ray data as compared to other $\HI$ maps with several uniform $\Ts$ from 125 K to 800 K.
Although gas properties in the cloud complex are discussed under the assumption of CO-dark $\Htwo$ (e.g., \citealt{Wolfire+10}; \citealt{Smith+14}) as a candidate of the dark gas, quantitatively estimate of the optically thick $\HI$ is not performed in their studies.
The assumption of a uniform $\Ts$ makes difficult to examine the $\HI$ gas with low $\Ts$ present in the CNM (e.g., \citealt{HeilesTroland03b}).

On the other hand, F14 and F15 have attempted to examine a total column density model as a function of $\taud$.
The $\NH$ model not relying on a uniform $\Ts$ allows accurate measurements of the $\HI$ gas.
F14 and F15 found that the $\NH$ model reproduces the scatter correlation in the $\taud$--$\WHI$ relationship, suggesting a large amount of the optically thick $\HI$ around molecular clouds.
\citet{Okamoto+17} demonstrated that the $\NH$ model is applicable in the Perseus molecular clouds and suggested the average $\HI$ optical depth is up to $\sim$0.9. 
Using the obtained $\NH$ model, the authors also derived a spatial distribution of the CO-to-$\Htwo$ conversion factor ($\XCO$) with the average value $\sim$1.0 $\times$ 10$^{20}$ $\UXCO$, which is comparable to past measurements of the Galactic interstellar clouds (e.g., \citealt{Bolatto+13}).

In this paper, we aim to investigate gas properties in the Chamaeleon region, focusing on the optically thick $\HI$, by attempting the method applied in \citet{Okamoto+17}.
The aim of the present paper is summarized below.

\begin{itemize}
\item Applying the $\taud$-based $\NH$ model to the Chameleon complex and to understand the physical states of the $\HI$ gas through a comparison with F14 (MBM 53, 54, 55 and HLC G92−35, hereafter denote as MBM~53--55), F15 ($|$$b$$|$ $>$ 15$^{\circ}$ in the all sky), and \citet{Okamoto+17} (Perseus) and derive the distribution of an $\XCO$ factor.
\item To test dust evolution found in the Orion A \citep{Roy+13} and the Perseus \citep{Okamoto+17} regions. This will bring a better understanding of dust properties in the local ISM.
\end{itemize}

This paper is organized as follows. 
Section 2 shows observational datasets.
Section 3 summarizes gas properties in the Chamaeleon region. 
Section 4 describes the $\NH$ model applied in this study.
In Section 5, we discuss possibility of the optically thick $\HI$ and these gas properties in comparison with other local molecular clouds.
A summary is given in Section 6.
All velocity information in the present paper is represented by local standard of rest (denoted as $\VLSR$).

\clearpage

\section{Observational Datasets} \label{sec:ObsDatasets}

To investigate gas and dust properties in the Chamaeleon region, we have used the following datasets. 

\subsection{$\HI$ Data} \label{sec:HIdata}

The Galactic All-Sky Survey (GASS) conducted with the Parkes 64 m radio telescope has provided the most sensitive and the highest resolution data of the $\HI$ 21 cm line emission for the southern sky (\citealt{McClure+09}; \citealt{Kalberla+10}; \citealt{KalberlaHaud15}). 
In this study, we have used the second released GASS data\footnote{The $\HI$ 4$\pi$ (HI4PI) survey \citep{Bekhti+16} adopting the third revision of GASS data \citep{KalberlaHaud15} were released in 2016. Instrumental effects remained in the past GASS data are corrected. We confirmed that our results do not change significantly even if we use these revised data.}, in which effects of stray radiation received by the antenna diagram have been corrected \citep{Kalberla+10}.
We have kept HPBW 16$\arcmin$ and the velocity resolution 0.82 km s$^{-1}$ in the original data.
A typical noise level for the analysis region in root-mean-square (rms) is $\sim$0.05 K per channel.
The measured velocity range for the Chamaeleon region is $-$500 km s$^{-1}$ to $+$400 km s$^{-1}$, but most of the $\HI$ line velocities span $-40$~km~s$^{-1}$~$\lesssim$~$\VLSR$~$\lesssim$~$+$20~km~s$^{-1}$ except for possible contribution from the Large Magellanic Cloud (LMC) at $+200$~km~s$^{-1}$~$\lesssim$~$\VLSR$~$\lesssim$~$+$300~km~s$^{-1}$.

\subsection{CO Data} \label{sec:COdata}
To trace the distribution of molecular hydrogen, we have used $^{12}$CO $J$$=$1--0 emission line observed by the NANTEN 4 m millimeter telescope located at Las Campanas, Chile.
NANTEN observations toward the Chamaeleon region were performed during two periods, from July to September in 1999 and from October to December in 2000 \citep{Mizuno+01}.
The HPBW of the data is 2$\farcm$6 at 115 GHz with grid spacing of 8$\arcmin$ and typical noise fluctuation is $\sim$0.3 K at a velocity resolution of 0.1~km~s$^{-1}$.

\subsection{{\it Planck} and {\it IRAS} Dust Emission Data}
The {\it IRAS} and {\it Planck} satellites performed the all-sky survey in millimeter/submillimeter wavelength, providing high-quality data of the thermal dust emission.
The measured intensities of the {\it Planck} 353, 545, and 857 GHz data and of the IRIS (Improved Reprocessing of the {\it IRAS} Survey) 100 $\mu$m data were fitted by modified-blackbody intensity spectra \citep{Planck11b}, which reveals dust properties  down to 5$\arcmin$ spatial resolution in HPBW with relative accuracy of $\sim$10\%.
In this analysis, we have used the dust optical depth at 353 GHz ($\taud$) and the dust temperature ($\Td$) to model/evaluate  the total gas column density.
The data with HEALPix format \citep{Gorski+05} released version R1.10\footnote{http://irsa.ipac.caltech.edu/data/Planck/release\_1/all-sky-maps/} are used.

\subsection{$J$-band Extinction Data}

Using the 2MASS (two micron all-sky survey) infrared extinction data measured at $J$, $H$, and $K$ bands, \citet{JuvelaMontillaud16} have derived interstellar extinction map at the $J$ band ($\AJ$) over the whole sky, with optimal techniques to map the dust column density (``NICER''; \citealt{LombardiAlves01} and ``NICEST''; \citealt{Lombardi09}).  
In the present study, we have used the $\AJ$ map constructed with the ``NICEST'' method to investigate correlation with the dust optical depth for the Chamaeleon complex. 
An all-sky map given in magnitude at the spatial resolution 3$\arcmin$ (full width half maximum) with the HEALPix format was downloaded from the archival page\footnote{http://www.interstellarmedium.org}.
The typical noise fluctuation for the Chamaeleon region is $\sim$0.08 mag in rms.

\subsection{$\Halpha$ Data}
We have used the optical $\Halpha$ data obtained by \citet{Finkbeiner03} in order to identify the bright $\HII$ regions, where dust grains are heated up or destroyed by ultraviolet (UV) radiation and the neutral hydrogen is ionized as well. 
In the present study, these regions are masked to avoid mixing different gas properties in local specific environment  (e.g., faint diffuse gas exposed by the strong radiation near the Galactic plane).
The typical sensitivity for the Chamaeleon region is estimated to be $\sim$0.3 R and the spatial resolution is 6$\arcmin$ in HPBW.

\subsection{21 cm Radio Continuum Data}
Radio continuum data have been used to estimate contribution from the background radiation, including the 2.7~K cosmic microwave background.
We have used the 21 cm (1.4 GHz) ``CHIPASS'' continuum map \citep{Calabretta+14}, which have been constructed by a combination of $\HI$ data obtained from the Parkes All-Sky Survey and Zone of Avoidance survey.
The typical sensitivity is $\sim$40 mK and the spatial resolution is 14$\farcm$4 in HPBW.

\section{Gas and Dust Properties in Chamaeleon Region} \label{sec:ChamRegion}

\subsection{Gas and Dust Spatial Distributions}

\begin{figure*}[h]
 \begin{center}
  \includegraphics[width=180mm]{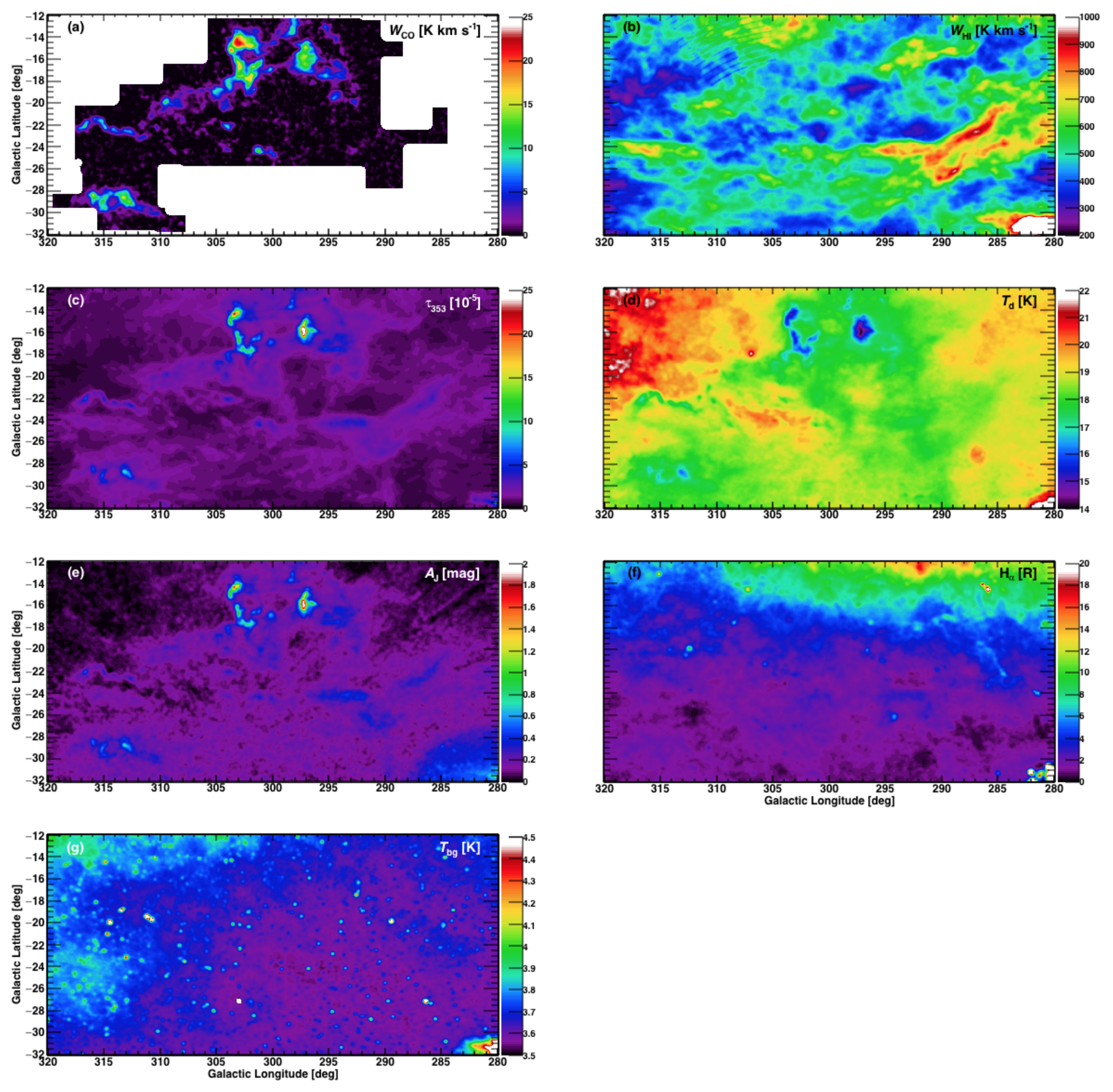}
  \end{center}
 \caption{Spatial distributions of the gas and dust propertis in the Chamaeleon region corrected for the 16$\arcmin$ effective beam size: (a) $\WCO$ obtained with the NANTEN telescope. (b) $\WHI$ from the GASS data. (c) $\taud$ and (d) $\Td$ based on the {\it Planck}/{\it IRAS} observations. (e) $\AJ$ obtained through the NICEST method using the 2MASS data. (f) $\Halpha$ intensity map from \citet{Finkbeiner03}. (g) 21 cm radio continuum map \citep{Calabretta+14}. Details are described in the text.} 
\label{fig:fig1}  
\end{figure*}

Using the datasets described in Section~\ref{sec:ObsDatasets}, we have made maps showing spatial distributions of the gas and dust properties in the Chamaeleon region, which are summarized in Figure~\ref{fig:fig1}.
Each map on the panels (a)--(h) are described below.

\begin{enumerate}
\renewcommand{\labelenumi}{(\alph{enumi})}

\item Velocity-integrated intensity map of the $^{12}$CO $J$$=$1--0 line (hereafter denoted as $\WCO$) obtained by the NANTEN 4 m telescope. The integrated velocity range is from $-$16 km s$^{-1}$ to $+$16 km s$^{-1}$, where most of the emission is included.
The peak intensity in $\WCO$ $\sim$25 K km s$^{-1}$ is intermediate between those measured from the MBM~53--55 \citep{Yamamoto+03} and Perseus \citep{Okamoto+17} molecular clouds.

\item Velocity-integrated intensity map of the $\HI$ 21~cm line ($\WHI$) obtained from the GASS data (\citealt{McClure+09}; \citealt{Kalberla+10}). 
The integrated velocity range is from $-$500 km s$^{-1}$ to $+$400 km s$^{-1}$ in the original data.
Although scanning effects are found in 300$^{\circ}$ $\lesssim$ $l$ $\lesssim$ 310$^{\circ}$, $-$20$^{\circ}$ $\lesssim$ $b$ $\lesssim$ $-$12$^{\circ}$ (see also Figure 8 in \citealt{KalberlaHaud15}), we confirmed that they do not affect the result of this study. 
The gas along the line of sight within the region is mainly separated into three components on the basis of the velocity line profile (see Section~\ref{sec:VelocityStructure}).

\item $\taud$ map obtained from the thermal dust emission model based on the {\it Planck/IRAS} data.

\item Dust temperature map obtained from the thermal dust emission model based on the {\it Planck/IRAS} data.

\item $J$-band extinction ($\AJ$) \citep{JuvelaMontillaud16} obtained with ``NICEST'' method \citep{LombardiAlves01} using the 2MASS near infrared data.

\item $\Halpha$ intensity map to search the $\HII$ regions.

\item Brightness temperature at 21 cm wavelength derived by \citet{Calabretta+14}. 
The map has been used to estimate the background brightness temperature at 21 cm ($\Tbg$) (see Section~\ref{sec:ModelGasColumnDensity}).

\end{enumerate}

\subsection{Masking Areas}

\begin{figure*}[h]
 \begin{center}
  \includegraphics[width=140mm]{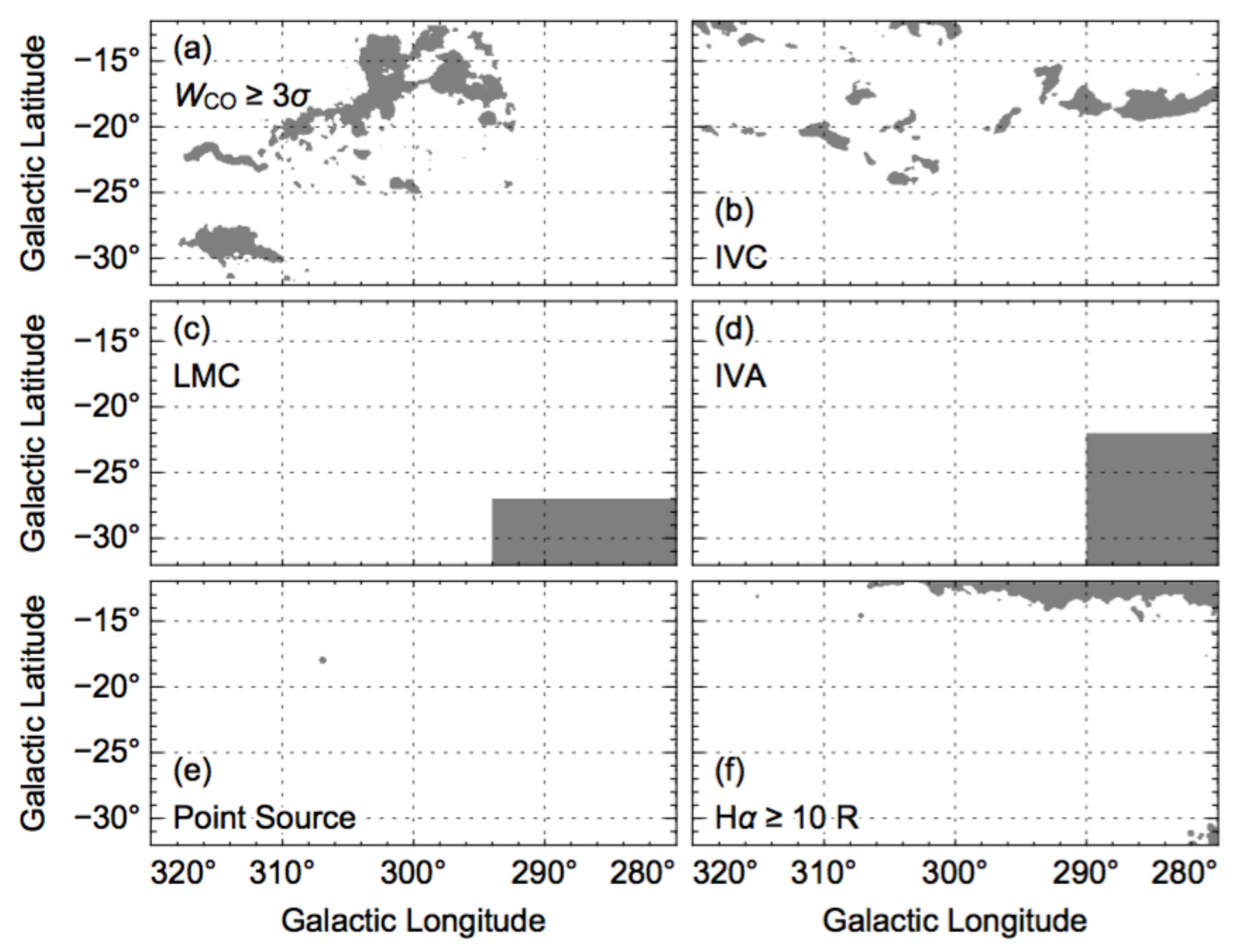}
  \end{center}
 \caption{Masked areas (shown by the shaded color) applied in the present analysis: (a) molecular gas with $\WCO >$ 3 $\sigma$.  (b) IVCs with $\WHI$ $>$ 50 K km s$^{-1}$ at $\VLSR$ $<$ $-$30 km s$^{-1}$. (c) LMC- and (d) IVA-dominated regions. (e) Position of a Be star, HIP 70248. (f) Ionized gas with $\Halpha$ $>$ 10 R. Details are described in the text.}
\label{fig:fig2}  
\end{figure*}

In the present study, when focusing on the atomic gas data, we have masked the molecular gas regions to remove data points with possible contamination from the high-density regions.
We also applied the mask to several areas having velocity profiles different from the local clouds with the peaks at $V_{\rm LSR}$ $\sim$ 0 km s$^{-1}$ (see also Section~\ref{sec:VelocityStructure}) and regions heated by the interstellar radiation field (ISRF), which may change the local gas-to-dust ratio.
The areas including the LMC components are also masked, since their gas-to-dust ratio are much different from the local ISM.
Figures~\ref{fig:fig2}(a)--(f) show the masked regions applied in this analysis.

\begin{enumerate}
\renewcommand{\labelenumi}{(\alph{enumi})}

\item The areas with significant CO emission ($\WCO$ $>$ 1.2 K km s$^{-1}$ ($\sim$3 $\sigma$)), where $\Htwo$ is dominant compared to $\HI$.

\item Intermediate-velocity clouds (IVCs) observed in the negative velocity range.
The areas with $\WHI$ $>$ 50 K km s$^{-1}$ at $V_{\rm LSR}$ $<$ $-$30 km s$^{-1}$ are masked.

\item The areas including outskirts and streams around the LMC, which are seen around the bottom-right part in the analysis region (see the gas and dust distributions in Figures~\ref{fig:fig1}(b)--(g)).

\item Intermediate-velocity arc (IVA) consisting of $\HI$-dominated clouds characterized by an elongated gas structure across the whole longitude direction at $-15$ km s$^{-1}$ $\lesssim$ $\VLSR$ $\lesssim$ $-$5 km s$^{-1}$ \citep{Planck15}. 
We masked areas where the contribution from IVA is more dominant ($l <$ 290$^{\circ}$ and $b <$ $-$22$^{\circ}$) compared to the  local $\HI$ clouds (see details in Section~\ref{sec:VelocityStructure}).

\item The position around a Be star HIP 70248 located at ($l$,$b$) $\sim$ (306$\fdg$9, $-$18$\fdg$0) 

\item The region with $\Halpha$ $>$ 10 R, where the dust grains are heated up or destroyed and hydrogen gas is ionized.

\end{enumerate}
\noindent
The mask (a) is applied to Figures~\ref{fig:fig5}(a), \ref{fig:fig7}, \ref{fig:fig9}(a), \ref{fig:fig10} (except for the $\NH$ histogram in the panel (c)), \ref{fig:fig14}(c) and \ref{fig:fig17}.
The other masks are applied to Figures~\ref{fig:fig4}--\ref{fig:fig11},  \ref{fig:fig13}, \ref{fig:fig14}(c) and  \ref{fig:fig18}.

\clearpage

\subsection{Velocity Structure in the Neutral Gas} \label{sec:VelocityStructure}

\begin{figure*}[h]
 \begin{tabular}{cc}
  \begin{minipage}{0.5\hsize}
   \begin{center}
    \rotatebox{0}{\resizebox{8cm}{!}{\includegraphics{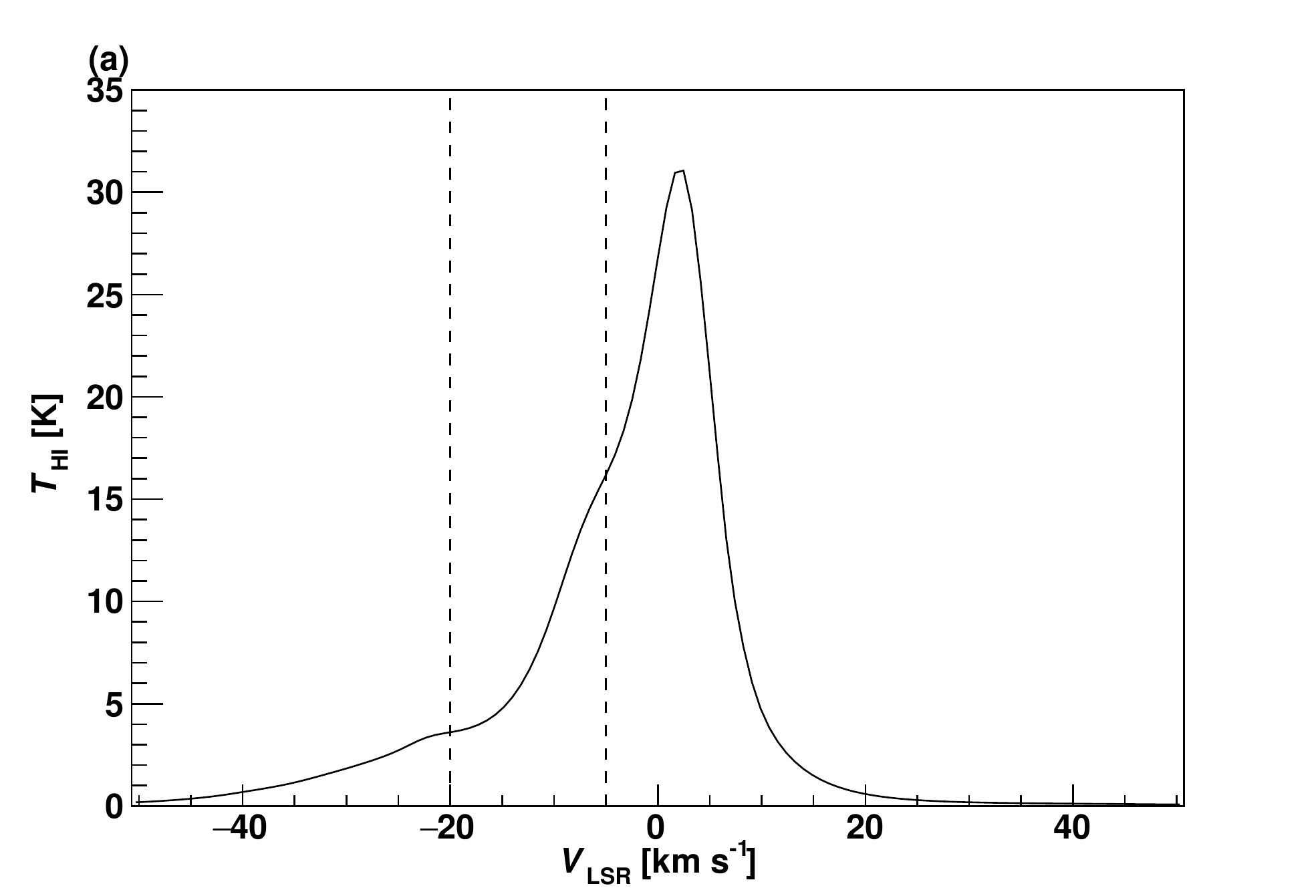}}}
   \end{center}
  \end{minipage} 
  \begin{minipage}{0.5\hsize}
   \begin{center}
    \rotatebox{0}{\resizebox{8cm}{!}{\includegraphics{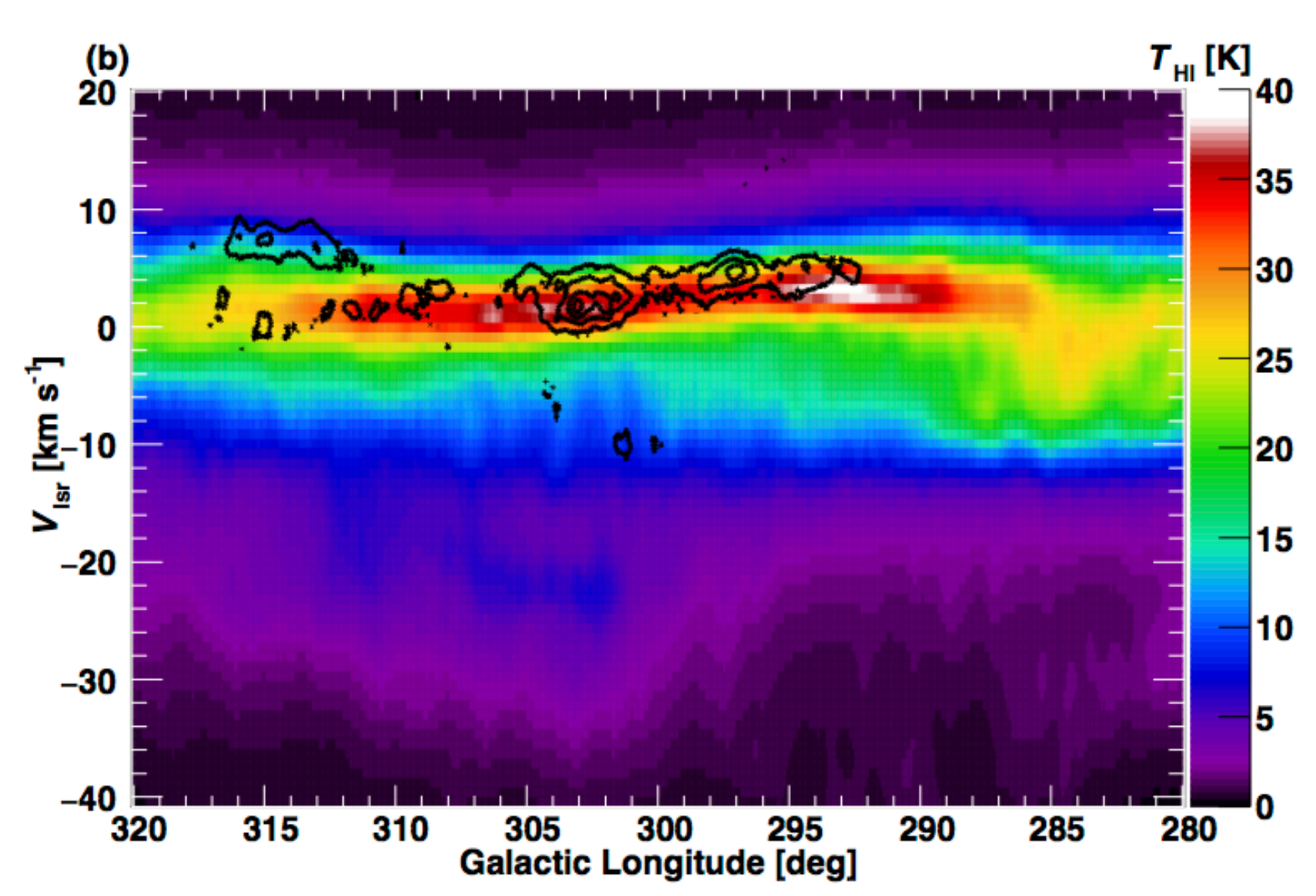}}}
   \end{center}
  \end{minipage} \\
  \end{tabular}  
  \caption{$\HI$ and $^{12}$CO $J$$=$1--0 velocity structures for the Chamaeleon region. (a) $\HI$ average spectrum giving the brightness temperature ($\THI$) of the region analyzed. The vertical dashed lines indicate velocity boundaries roughly separated by eyes. Examples of the line profiles having the strong emission from the local, IVA and the high velocity components are shown in Figure~\ref{fig:fig16}. (b) Longitude-velocity diagram of the $\HI$ (image) and CO (contours) represented by the average intensities within the latitude range $-32^{\circ} \le b \le -12^{\circ}$. The CO contours are drawn every 0.25~K step from the minimum intensity 0.03~K ($\sim$3 $\sigma$).}
\label{fig:fig3}   
\end{figure*}

Line profiles obtained from the $\HI$ and CO data allow kinematical separations of the gas distribution.
Figure~\ref{fig:fig3}(a) indicates an $\HI$ intensity spectrum showing the brightness temperature ($\THI$) averaged in the analysis region.
Figure~\ref{fig:fig3}(b) is an average longitude-velocity diagram, which overlays the CO intensity represented by the black contours.
We have found three components separated by the dashed vertical lines in the spectrum, whose gas structures are seen in the longitude-velocity diagram and a velocity channel map shown in Figure~\ref{fig:fig15}: (i) local clouds with the intensity peak around $\VLSR=$~0~km~s$^{-1}$, at which most of the CO emission is detected; (ii) wing-like structure at the intermediate velocity range spanning $-15$~km~s$^{-1}$~$\lesssim \VLSR \lesssim-5$~km~s$^{-1}$, which corresponds to the IVA \citep{Planck15} characterized by a mild gradient in the velocity toward the negative longitude direction, with comparatively bright $\HI$ emission at 280$^{\circ}$~$\lesssim l \lesssim$~290$^{\circ}$; (iii) high velocity component with a long tail at $\VLSR \lesssim$~$-20$~km~s$^{-1}$, which corresponds to faint diffuse emission observed at 302$^{\circ}$~$\lesssim l \lesssim$~314$^{\circ}$, exhibiting bridging feature connected to the intermediate clouds.
To more clearly show that the $\HI$ spectrum can be separated into the three components, we give examples of the spectra in Figure~\ref{fig:fig16} for restricted regions with the size of 1$\fdg$0 $\times$ 1$\fdg$0.
\citet{Planck15} also shows that the similar spectral separation can be applied in the Chamaeleon region.
In order to remove contamination from clouds other than the local Chamaeleon complex, we have masked the region at (280$^{\circ}$~$\leq l \leq$~290$^{\circ}$, $b \leq -22$$^{\circ}$), where contribution from the IVA is relatively large, and the regions with strong emission from the high velocity component.  
Although the IVA component is extended toward $l \gtrsim$~290$^{\circ}$, from which the faint diffuse emission at the high velocity range is observed, these contributions are not significant compared to that from the local $\HI$ clouds lying at the same line of sight.
We therefore do not mask this area. 

\subsection{Correlation between Gas and Dust Properties} \label{sec: CorrelationGasDust}

We have investigated gas and dust properties using measurements of the {\it Planck} dust emission model and the near infrared extinction, $\AJ$. 
The $J$-band wavelength ($\sim$1.25 $\mu$m) is much larger than a typical size of the dust particle ($\lesssim$~0.2 $\mu$m; \citealt{Jones+13}) even if we take into account changing the size of a dust particle in its evolution (within $\sim$0.02 $\mu$m; \citealt{Ysard+15}).
\citet{Martin+12} suggests that the ratios of near-infrared color excess to $\NH$ change less significantly in evolution process of the dust grains.
These results expect that the ratio of $\AJ$/$\NH$ does not change significantly, indicating that $\AJ$ can be a tracer to measure the total gas column density.

\begin{figure*}[h]
 \begin{center}
  \includegraphics[width=90mm]{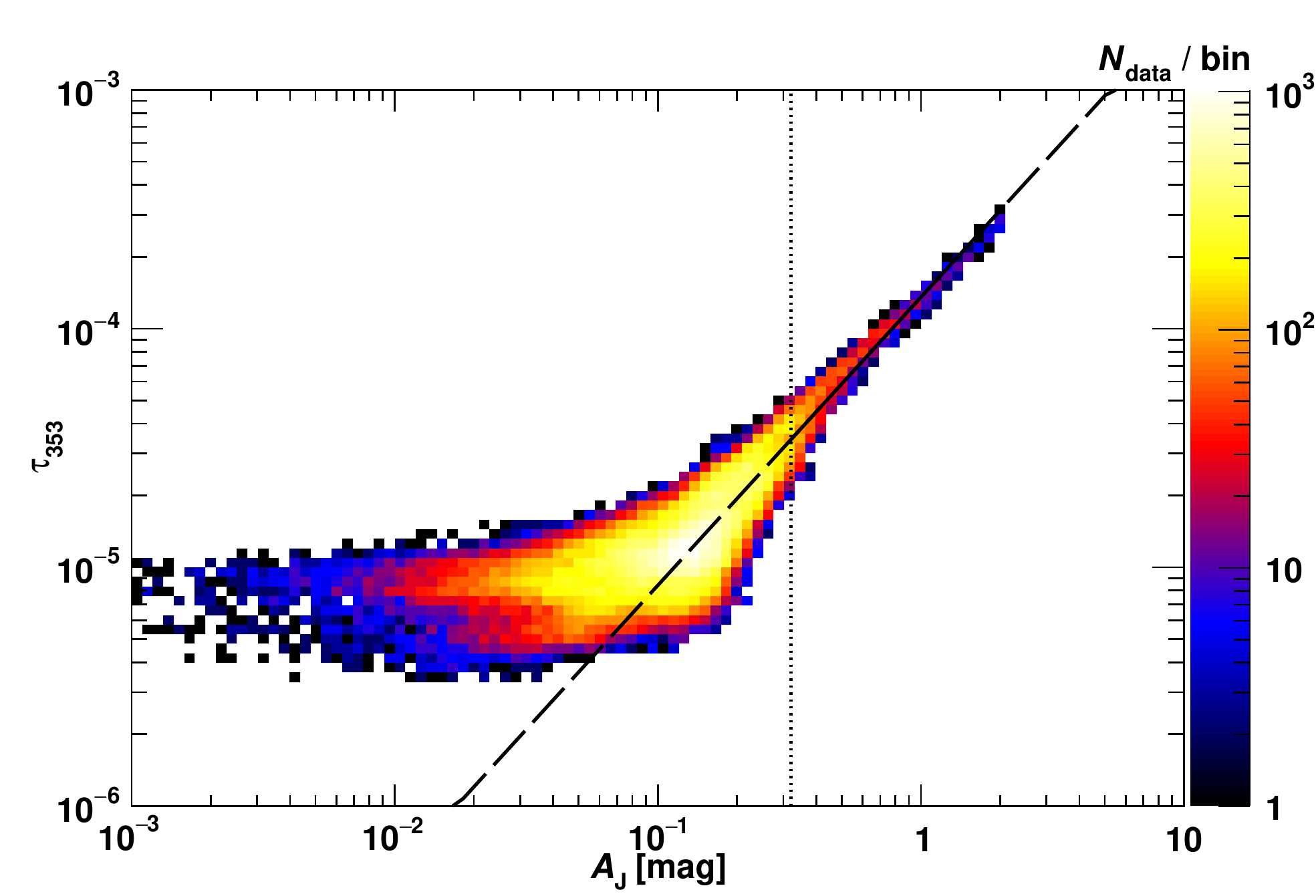}
  \end{center}
 \caption{Correlation between $\AJ$ and $\taud$. The dashed line indicates the best-fit liner regression, $\taud \propto \AJ^{1.21\pm0.04}$ obtained by the fit to the data points above 0.32 mag ($\sim$4~$\sigma$) in $\AJ$ shown by the vertical dotted line.}
\label{fig:fig4}  
\end{figure*}

Figure~\ref{fig:fig4} shows a correlation plot between $\taud$ and $\AJ$ for the Chamaeleon region.
Whereas most of the data are saturated in low $\AJ$, data above a few 0.1 mag show a tight correlation with $\taud$.
The correlation between the two quantities above 0.32 mag ($\sim$4~$\sigma$ in $\AJ$) is expressed by a regression line as follows,

\begin{eqnarray}
\taud = \left[\left(1.35 \pm 0.05 \right) \times 10^{-4} \right] \times \AJ^{1.21 \pm 0.04},
\label{eq:tau353_AJ} 
\end{eqnarray}

\noindent
which is shown by the dashed line in the figure.
This correlation between $\taud$ and $\AJ$ indicates that $\taud$ increases as equivalent to $\sim$1.2th power of $\NH$.
Similar studies for the Orion A cloud \citep{Roy+13} and the Perseus cloud \citep{Okamoto+17} found that nonlinear relations between dust optical depth and $\NH$, whose power-law exponents are 1.28$\pm$0.01 and 1.32$\pm$0.04, respectively.
These authors make a point that the nonlinear relation is due to dust evolution effects.
Although the correlation between gas and dust properties exhibits variations among the different regions, dust growth in the Chamaeleon region is also traced as the similar nonlinear relation with a power-law exponent of $\sim$1.2.

\begin{figure*}[h]
 \begin{tabular}{cc}
  \begin{minipage}{0.5\hsize}
   \begin{center}
    \rotatebox{0}{\resizebox{9cm}{!}{\includegraphics{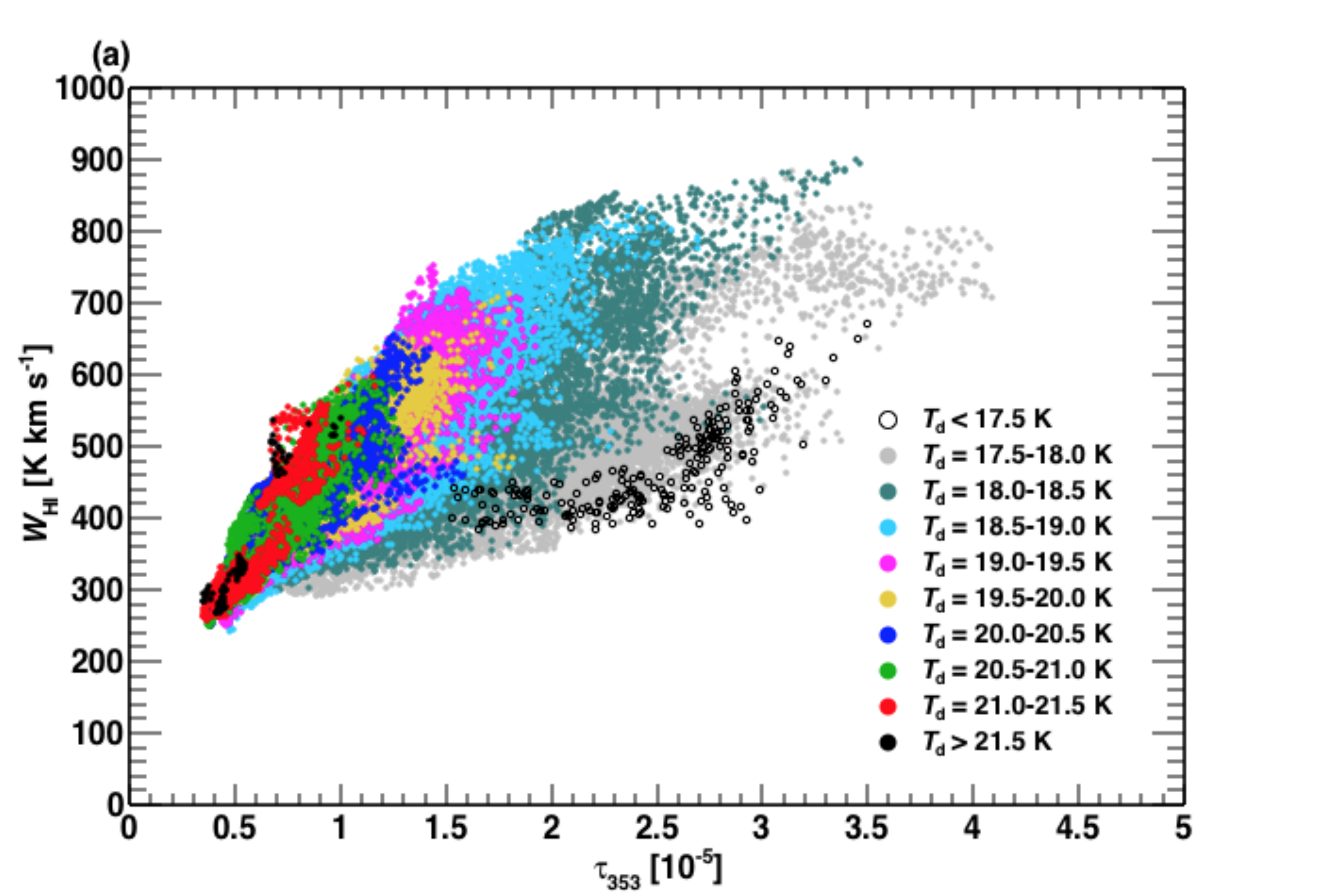}}}
   \end{center}
  \end{minipage} 
  \begin{minipage}{0.5\hsize}
   \begin{center}
    \rotatebox{0}{\resizebox{9cm}{!}{\includegraphics{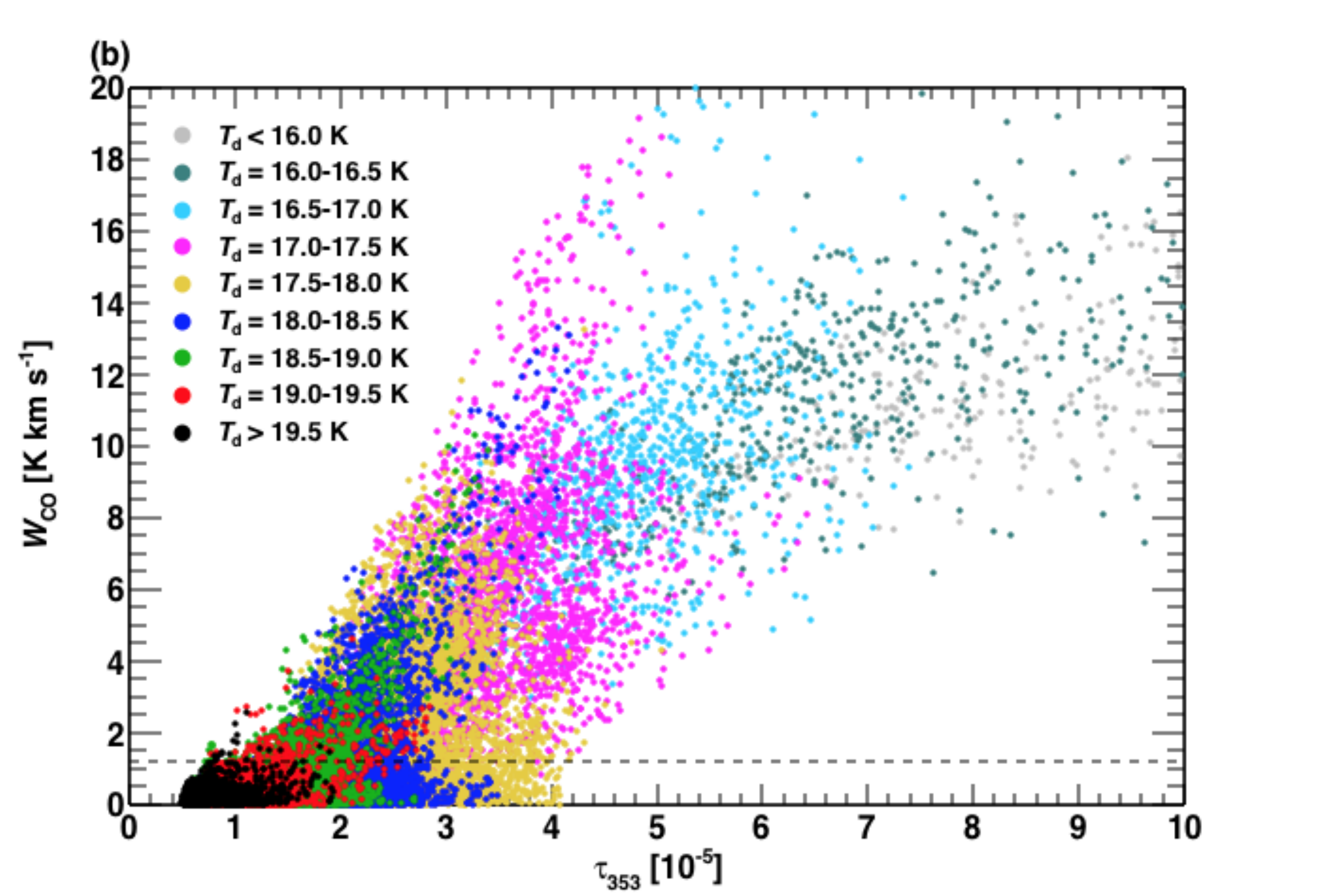}}}
   \end{center}
  \end{minipage} \\
  \end{tabular}  
  \caption{Correlation plots between (a) $\taud$ and $\WHI$ and (b) $\taud$ and $\WCO$, sorted by several $\Td$ intervals. The horizontal dashed line in the panel (b) indicates 3 $\sigma$ confidence level in $\WCO$. The same plots shown in the different panels sorted by $\Td$ are given in Figures~\ref{fig:fig17} ($\taud$--$\WHI$) and \ref{fig:fig18} ($\taud$--$\WCO$).}
\label{fig:fig5}   
\end{figure*}

\begin{figure*}[h]
 \begin{center}
  \includegraphics[width=90mm]{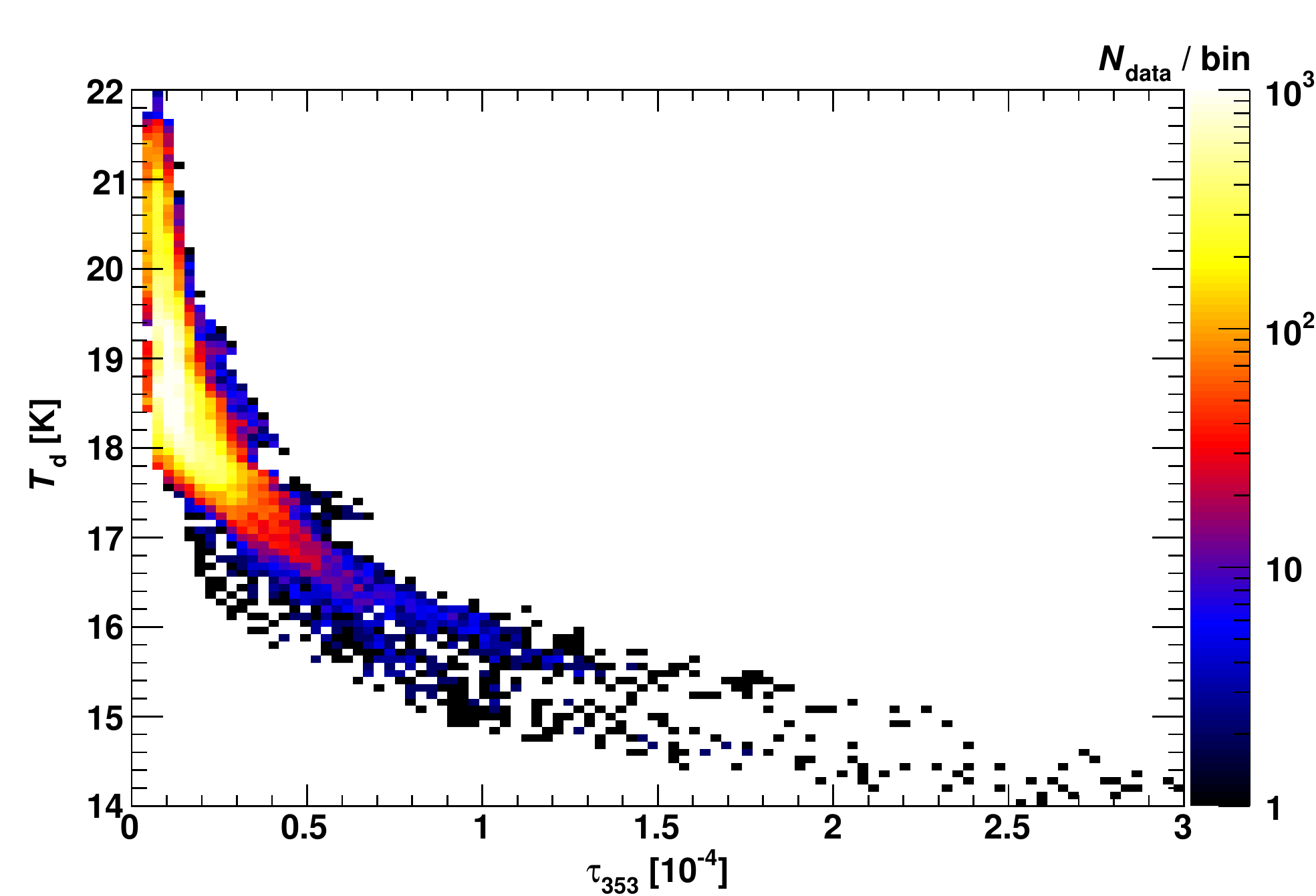}
  \end{center}
 \caption{Correlation between $\taud$ and $\Td$.}
\label{fig:fig6}  
\end{figure*}

Figures~\ref{fig:fig5}(a) and (b) show correlation plots of $\taud$ with $\WHI$ and $\WCO$, respectively, which are sorted by several $\Td$ in 0.5 K intervals represented by different colors.
To more clearly show the data points, the same density distribution plotted on different panels sorted by $\Td$ are given in Figures~\ref{fig:fig17} ($\taud$--$\WHI$) and \ref{fig:fig18} ($\taud$--$\WCO$).
Although the correlation between $\taud$ and $\WHI$ is poor overall,  
the data points sorted by $\Td$ exhibit clearly different correlations depending on $\Td$: 
the scattering is small in high $\Td$ areas and it becomes larger with decreasing $\Td$. 
The relationship with $\WCO$ does not show a tight correlation either, even if we exclude noisy signals in $\WCO$ below 1.2~K~km~s$^{-1}$ ($\sim$3 $\sigma$) and saturated ones above $\sim$8~K~km~s$^{-1}$.
However, contrast in the dispersion among the different $\Td$ is not large compared to the relation with $\WHI$. 
As shown in Figure~\ref{fig:fig6}, we have found an anti-correlation between $\taud$ and $\Td$ probably related to feedback from the ISRF: in low density regions with lower $\taud$, the ISRF heats up dust grains and leads to higher $\Td$. Conversely, in high density regions where $\taud$ is large, the ISRF is shielded by dust grains, which leads to lower $\Td$. 
Similar gas and dust properties are also found in other local molecular cloud complexes such as MBM~53--55 (F14) or Perseus \citep{Okamoto+17} regions.

\clearpage

\section{Modeling the Gas Column Density} \label{sec:ModelGasColumnDensity} 

The $\HI$ column density usually adopted as the optically thin limit ($\NHIstar$) is calculated from $\WHI$ as follows, 
\begin{eqnarray}
\NHIstar = \XHI \times \WHI,
\label{eq:NHImodelthin} 
\end{eqnarray} 
where $\XHI = 1.823 \times 10^{18}$ cm$^{-2}$ K$^{-1}$ km$^{-1}$ s.
If the $\HI$ optical depth, $\tauHI$, does not satisfy $\tauHI \ll$ 1, this approximation underestimates the true $\HI$ column density.
F14 and F15 adopted $\taud$ as an accurate tracer of $\NHI$ and suggested possible large amount of the optically thick $\HI$ in the local ISM.
F14 analyzed a molecular cloud region, MBM~53--55, whose gas density and star forming activities are relatively lower compared to other local molecular cloud regions (e.g., \citealt{Yamamoto+03}).
The tight correlation between $\taud$ and $\WHI$ in high $\Td$ areas can be approximated well by a liner regression with small dispersion.
The best-fit linear line is applied to the $\HI$ column density in the optically thin case (see Figure 3 in F14).
Subsequently, F15 examined an $\NH$ model with nonlinear relation with $\taud$ to consider dust evolution effect suggested by \citet{Roy+13}.
The model is expressed as,

\begin{eqnarray}
\left( \frac{\taud}{\taudref} \right) = \left( \frac{\NH}{\NHref} \right)^{\alpha}
\label{eq:NHmodel} 
\end{eqnarray}

\noindent
where $\taudref$ and $\NHref$ are reference points satisfying the relationship $\NHref = (1.15\times10^{8})\times \XHI \times \taudref$ (Equation (2) in F15).
\citet{Okamoto+17} applied the nonlinear $\NH$ model to further investigate gas properties in Perseus molecular clouds and revealed that the $\NH$ model with the $\sim$1.3th power more traces accurately the gaseous components from the diffuse medium to dense cores in the cloud complex.

Following the above studies, we have applied the nonlinear $\NH$ model and examined a possibility of the optically thick $\HI$ in the Chamaeleon region.    
Based on an assumption of uniform gas-to-dust ratio in the local ISM, we have adopted the power-law index, $\alpha =$~1.2, which is derived from the $\taud$--$\AJ$ relationship found in the opaque region (see Section~\ref{sec: CorrelationGasDust}).
Although this gas-to-dust relation is not obtained from the diffuse $\HI$ medium, the similar nonlinearity is confirmed in \cite{Roy+13} down to $\NH$$\sim$1$\times$10$^{21}$ cm$^{-2}$, which corresponds to the typical $\NH$ discussed in F15 and \cite{Okamoto+17}. 
If we take into account that the gas column density significantly affects the dust properties, this assumption is compatible as a first approximation to consider the dust evolution effect.
The reference points in the $\NH$ model are determined to be $\taudref = 7.8\times10^{-7}$ and $\NHref = 1.6\times10^{20}$ cm$^{-2}$, with an analytical method described in Appendix~\ref{sec:RefPoints}.
The $\NH$ model is thus expressed as, 

\begin{eqnarray}
\NH = \left(\frac{\taud}{\taudref}\right)^{1/\alpha}\times\NHref = \left(2.0\times10^{25}\right)\times \taud^{1/1.2}.
\label{eq:NHmodelCham} 
\end{eqnarray}

\noindent
Using the $\NH$ model, we have modeled the scatter correlation between $\taud$ and $\WHI$ with the following independent two equations (e.g., \citealt{DickeyLockman90}; \citealt{Draine11}): radiative transfer equation of the $\HI$ 21 cm line emission, 

\begin{eqnarray}
\WHI = (\Ts - \Tbg) \times \Delta \VHI \times \{1- {\rm exp} (-\tauHI) \},
\label{eq:RadTransfer} 
\end{eqnarray}
and the $\HI$ optical depth derived from the $\HI$ spin flip transition, 

\begin{eqnarray}
\tauHI = \frac{\NHI}{\XHI} \times \frac{1}{\Ts} \times \frac{1}{\Delta \VHI },
\label{eq:SpinFlipTrans} 
\end{eqnarray}
where $\Delta\VHI$ is spectral width in velocity, which can be defined as $\WHI/T_{\rm \HI}$.

\begin{figure}[h]
 \begin{center}
  \includegraphics[width=90mm]{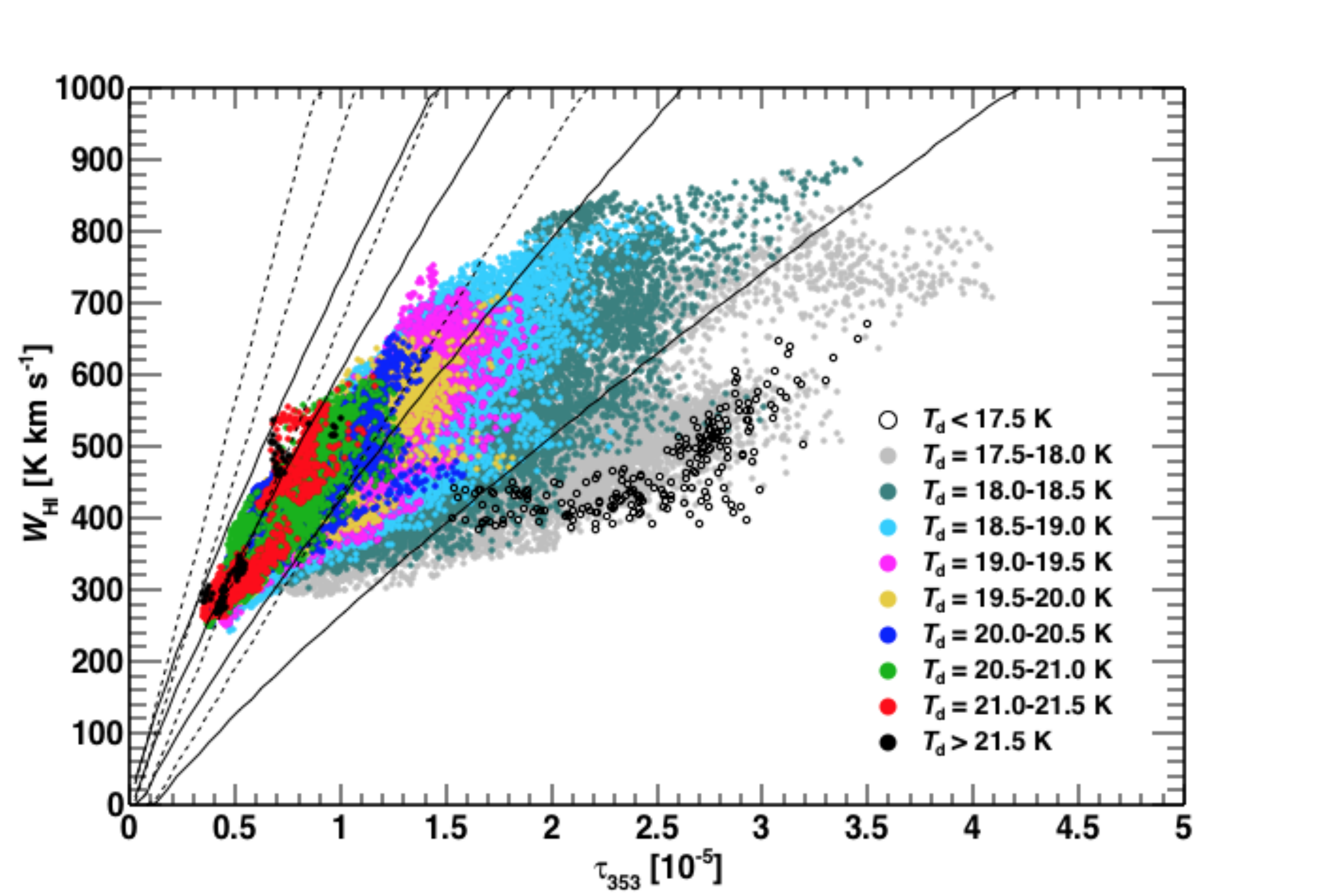}
  \end{center}
 \caption{Correlation between $\taud$ and $\WHI$ colored by $\Td$ in 0.5~K windows. The solid and dotted curves show the theoretical functions of Equation~(\ref{eq:WHIModelCurve}) for $\alpha =$~1.2 and 1.0, respectively, with $\tauHI \ll$ 1, $\tauHI =$ 0.34, 1.0 and 2.0 from left to right. The same plots shown in different panels of each $\Td$ are given in Figure~\ref{fig:fig17}.}
\label{fig:fig7}  
\end{figure}

Although $\Ts$ on the line of sight is not uniform, it can be approximated to a single component with a density-weighted harmonic mean in the line of sight (e.g., \citealt{Dickey+79}; \citealt{HeilesTroland03b}; \citealt{Fukui+18}).
By applying the $\NH$ model in Equation~(\ref{eq:NHmodelCham}) to the $\HI$-dominated region (i.e., $\NH = \NHI$), a coupled equation between Equations (\ref{eq:RadTransfer}) and (\ref{eq:SpinFlipTrans}) gives a theoretical model curve of $\WHI$ as a function of $\taud$, 

\begin{eqnarray}
\WHI = \left( \left(\frac{\taud}{\taudref}\right)^{1/\alpha} \times \frac{\NHref}{\XHI} \times  \frac{1}{\tauHI} \times \frac{1}{\Delta V_{\HI}} - T_{\rm {bg}} \right) \times \Delta V_{\HI} \times \{1 - {\rm exp}(- \tauHI)\}.
\label{eq:WHIModelCurve} 
\end{eqnarray}
Applying average values in the analysis region, $\Delta \VHI =$13 km s$^{-1}$ and $\Tbg =$ 3.7~K, which is estimated from the 21 cm radio continuum data, model curves with $\tauHI \ll$ 1, $\tauHI =$ 0.34, 1.0 and 2.0 are represented by the solid lines (from left to right) in Figure~\ref{fig:fig7}.
For comparison, model curves with $\alpha =$ 1.0 for the same $\tauHI$ are also shown.
The model curves with high $\tauHI$ allows to reproduce the scatter correlation in high $\taud$ areas. 
The model with the nonlinear relation, $\alpha =$ 1.2, traces better the mildly curved shape than $\alpha =$ 1.0.
This indicates that the $\NH$ model in Equation~(\ref{eq:NHmodelCham}) can approximately reproduce the gas distribution in the $\HI$-dominated region.
One can see a trend that $\tauHI$ becomes larger with decreasing $\Td$, which indicates that there is a positive correlation between $\Td$ and $\Ts$.
The contrast of the correlation strength among the different $\Td$ in the $\taud$--$\WCO$ relationship (Figures~\ref{fig:fig5}(b) and~\ref{fig:fig18}) is not significant  compared to the $\taud$--$\WHI$ relationship (Figures~\ref{fig:fig5}(a) and~\ref{fig:fig17}).
The different correlation strength in the $\taud$--$\WHI$ relationship among $\Td$ suggests existence of the atomic gas with low $\Ts$ (optically thick $\HI$) around the molecular clouds. 

\clearpage

\section{Discussion} \label{sec:discussion}

\subsection{Constituent of the Dark Gas}
\label{sec: DG}

Our study of the Chamaeleon region showed that the optically thick $\HI$ is important around the molecular clouds. 
However, the previous $\gamma$-ray study \citep{Planck15} disagrees with the large amount of the optically thick $\HI$. 
We discuss the cause of the contradiction below. 

First, it is to be recognized that the $\gamma$-ray study assumes a uniform and a high $\HI$ spin temperature above 100~K, which is a strong assumption that pre-excludes the high $\HI$ optical depth.
It is already shown by $\HI$ emission-absorption measurements \citep{HeilesTroland03b} that $\Ts$, the harmonic mean $\Ts$ in a line of sight, is as low as 40 K and that low $\Ts$ is appreciable in the CNM.
Further, $\Ts$ varies significantly from 30 K to more than 500 K, indicating that uniform high $\Ts$ is not supported. 
For realistic low $\Ts$ less than 100~K, $\HI$ optical depth is high, more than $\sim$1.0; 
$\tauHI$ derived from Equation~(\ref{eq:SpinFlipTrans})  is $\sim$1.1 at $\NHI =$~1.0$\times$10$^{21}$~cm$^{-2}$, close to the peak of $\NHI$ distribution, and $\Delta\VHI =$~5 km~s$^{-1}$, a typical line width of the CNM (c.f., F15). 
We note if $\tauHI$ is greater than 0.3 and 0.5, the optical depth correction by a factor of 1.2 and 1.3, respectively, is required in calculating $\NH$. 
In this sense, the boundary between optically thick or thin $\HI$ lines at $\tauHI$ of 0.3--0.5, depending on the accuracy needed, and the optically thin case is for $\tauHI$ less than 0.2 in the practical $\HI$ sensitivity. 
We expect that optically thick $\HI$ is suggested by a $\gamma$-ray study if the unrealistic assumption of the uniform high $\Ts$ is not adopted. 

It is discussed in the literature both observational and theoretical that there exists a significant amount of CO-dark H$_2$ gas (e.g., \citealt{Wolfire+10}; \citealt{Liszt+18}). 
We discuss critically these works below, and look into the cause of the apparent discrepancy.

\begin{itemize}
\item

Observations of molecular abundance are often used to estimate molecular fraction in the low density molecular gas.
Since absorption measurements toward radio continuum sources like quasars are more sensitive than emission line measurements, compact continuum sources are used to measure molecular absorption of HCO$^{+}$, HCN etc (e.g., \citealt{LisztGerin16}; \citealt{Liszt+18}; \citealt{Gerin+19}).
Detection of such molecular absorption shows that such rare molecules do exist in the low density gas where the gas may not be detectable in CO emission due to low density. 
\citet{Liszt+18} assumed that abundance ratio between HCO$^{+}$ and H$_{2}$ is uniform at 3$\times$10$^{-9}$ to derive H$_2$ density and argued that CO-dark H$_{2}$ gas may contribute significantly as the dark gas.
These authors derived high CO intensity by assuming the $\XCO = 2\times$10$^{20}$ cm$^{-2}$ K$^{-1}$ km$^{-1 }$s, which is about 2 times higher than those obtained by recent studies of the Chamaeleon region (\citealt{Ackermann+12}; \citealt{Planck15}), and showed that the predicted CO intensity disagrees with the non-detection of CO by NANTEN by a factor of more than 2 at several points in these clouds.
It is probable that their method is not accurate in the order of 10\%, because the molecular abundance varies significantly from place to place; \citet{Gerin+19} showed that HCO$^{+}$ abundance by ALMA observations, for instance, varies by an order of magnitude for density around $\NH$ of 10$^{20}$ cm$^{-2}$ (see their figure 4), and the assumption of uniform HCO$^{+}$ abundance by \citet{Liszt+18} is not supported.
The large discrepancy above in the expected CO intensity is explained as due to the unrealistic assumption of uniform HCO$^{+}$ abundance. 
In summary, the HCO$^{+}$ absorption is not an accurate method to calculate H$_2$ abundance because of the small and variable abundance, and the mass estimate of CO-dark gas is uncertainty by a factor of 2--3 at best. 
Accordingly, the molecular absorption is not accurate enough to constrain the dark molecular gas.

\item

\citet{Wolfire+10} made a theoretical study of CO-dark H$_2$ by using calculations of molecular abundance in a model cloud, and concluded that CO-dark gas is significant. 
This study assumes as the initial condition that the gas density is as high as 10$^3$ cm$^{-3}$, where hydrogen exists mostly as H$_2$.
It is a question if the initial condition is justified. 
\citet{InoueInutsuka12} showed how molecular gas is formed in the interstellar space by taking $\HI$ gas as the initial condition. 
This is a more general assumption and their results show that $\HI$ gas remains significant even after the convergence of $\HI$ flows after 10 Myrs.
The results suggest a mixture of $\HI$ and H$_2$ gas as the state of the realistic interstellar molecular gas.
Therefore, the fraction of H$_2$ is highly dependent on the initial condition. 
Since molecular gas is formed from atomic gas, any simulations assuming pure H$_2$ as the initial condition needs justification before confronting with observations.  
\end{itemize}

\citet{Fukui+18} performed a synthetic observation of the interstellar gas based on the results of \citet{InoueInutsuka12}. 
They applied molecular fraction measured by UV absorption \citep{Gillmon+06} to that of the simulated interstellar gas, whose $\NH$ is peaked at $\sim$1 $\times$ 10$^{21}$~cm$^{-2}$ in a range from 5$\times$10$^{20}$~cm$^{-2}$ to 2$\times$10$^{21}$~cm$^{-2}$, which is consistent wit the $\NH$ distribution obtained by the $\taud$-based $\NH$ model (see figure 4 in \citealt{Fukui+18}).
The results showed that the CNM has clumpy gas distribution with volume filling factor ($\sim$4\%) and gas density (10$^2$--10$^3$ cm$^{-2}$), wihch are consistent with generally suggested gas properties of the ISM; the gas masses of the CNM and WNM are comparable, while their density fraction is estimated to be 30:1 and thus the ratio of the volume filling factor is 1:30 (e.g., \citealt{DopitaSutherland03}).
The fraction of the CNM in $\int \tauHI dv$ does not contradict the $\HI$ emission-absorption measurements by \citet{HeilesTroland03a} (see Figure 13 in \citealt{Fukui+18}). 
The simulation showed that the fraction of H$_2$ is only $\sim$5\%. 

From the above discussions, we conclude that in the Chamaeleon region optically thick $\HI$ is more important than CO-dark H$_2$ and thus a dominant constituent of the dark gas. 
Hereafter we discuss gas properties and distribution, focusing on the optically thick $\HI$.

\subsection{Gas Distribution and Mass Fraction}
\label{sec: GasDistMassFrac}

\begin{figure*}[h]
 \begin{center}
  \includegraphics[width=140mm]{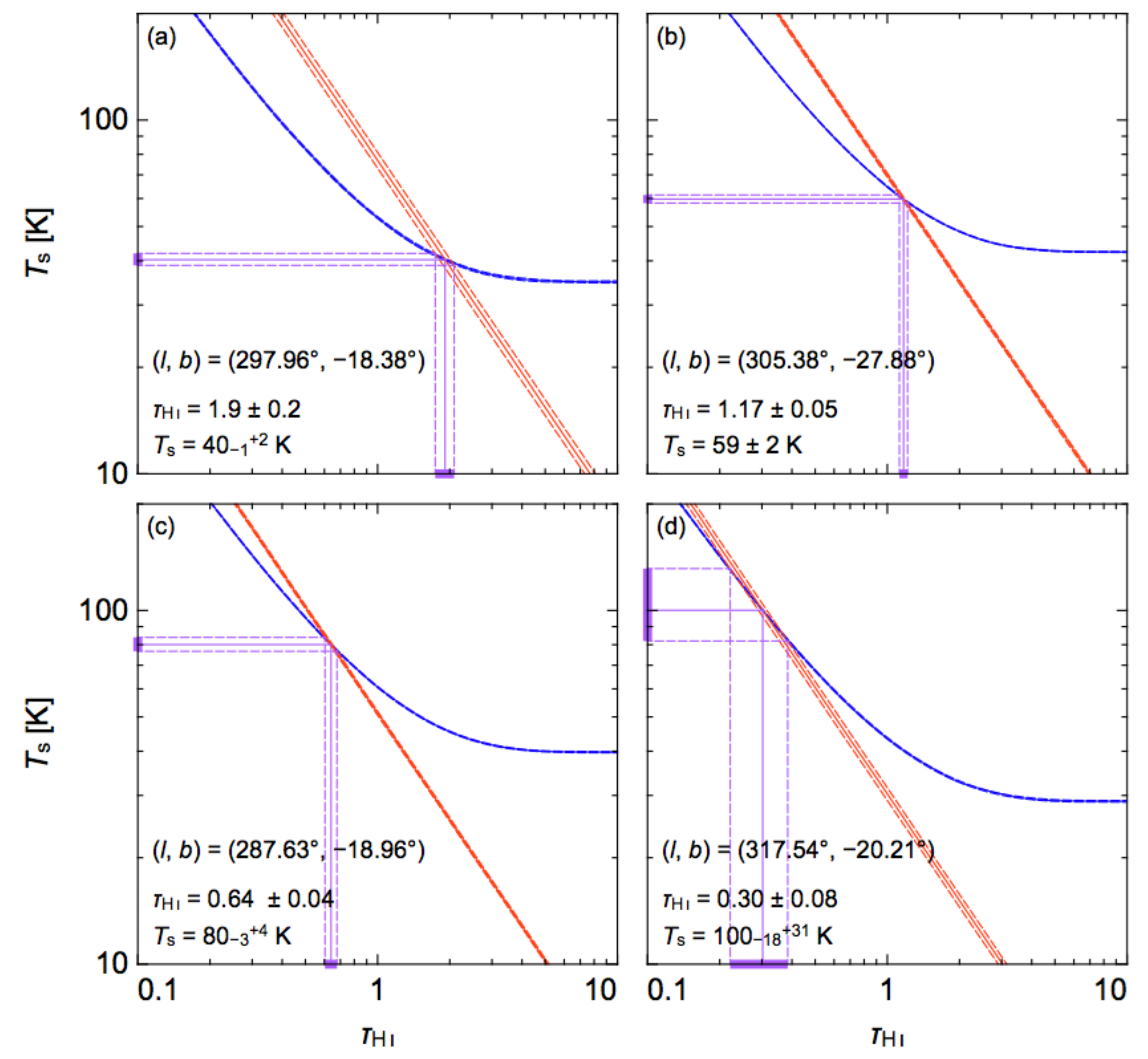}
  \end{center}
 \caption{Examples of the $\tauHI$ and $\Ts$ calculations. The blue and red solid lines indicate Equations (\ref{eq:RadTransfer}) and (\ref{eq:SpinFlipTrans}), respectively. The purple solid lines show the solution of $\tauHI$ and $\Ts$. The dashed lines of each color indicate 1 $\sigma$ error limit.}
\label{fig:fig8}  
\end{figure*}

With the $\NH$ model in Equation (\ref{eq:NHmodelCham}), we discuss $\HI$ gas properties and mass fractions in the  cloud complex.
We have solved the coupled equations between Equations (\ref{eq:RadTransfer}) and (\ref{eq:SpinFlipTrans}) and derived $\tauHI$ and $\Ts$ values by numerical solution.
Figures~\ref{fig:fig8}(a)--(d) show examples of the solution for the corresponding positions. 
The blue and red lines indicate Equations (\ref{eq:RadTransfer}) and (\ref{eq:SpinFlipTrans}), respectively, and the obtained $\tauHI$ and $\Ts$ are shown by the purple solid line.
The dashed lines indicate 1$\sigma$ error limits. 
The errors in the obtained $\tauHI$ and $\Ts$ are given by the dashed crossing points.
These examples give solutions ($\tauHI$, $\Ts$ (K)) $\sim$ (1.9, 40), (1.2, 60), (0.6, 80), and (0.3, 100).

\begin{figure*}[h]
 \begin{tabular}{cc}
  \begin{minipage}{0.5\hsize}
   \begin{center}
    \rotatebox{0}{\resizebox{9cm}{!}{\includegraphics{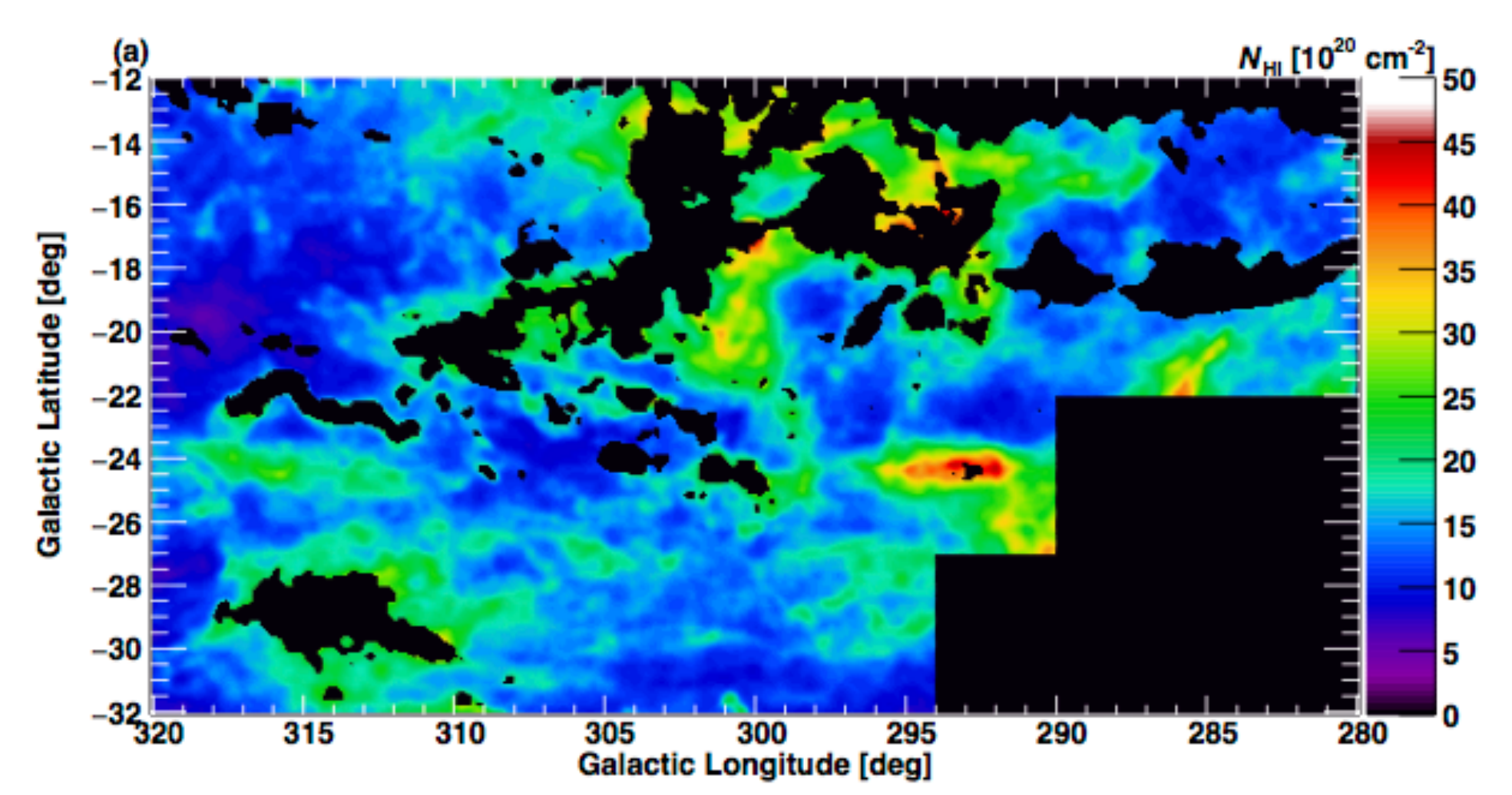}}}
   \end{center}
  \end{minipage} 
  \begin{minipage}{0.5\hsize}
   \begin{center}
    \rotatebox{0}{\resizebox{9cm}{!}{\includegraphics{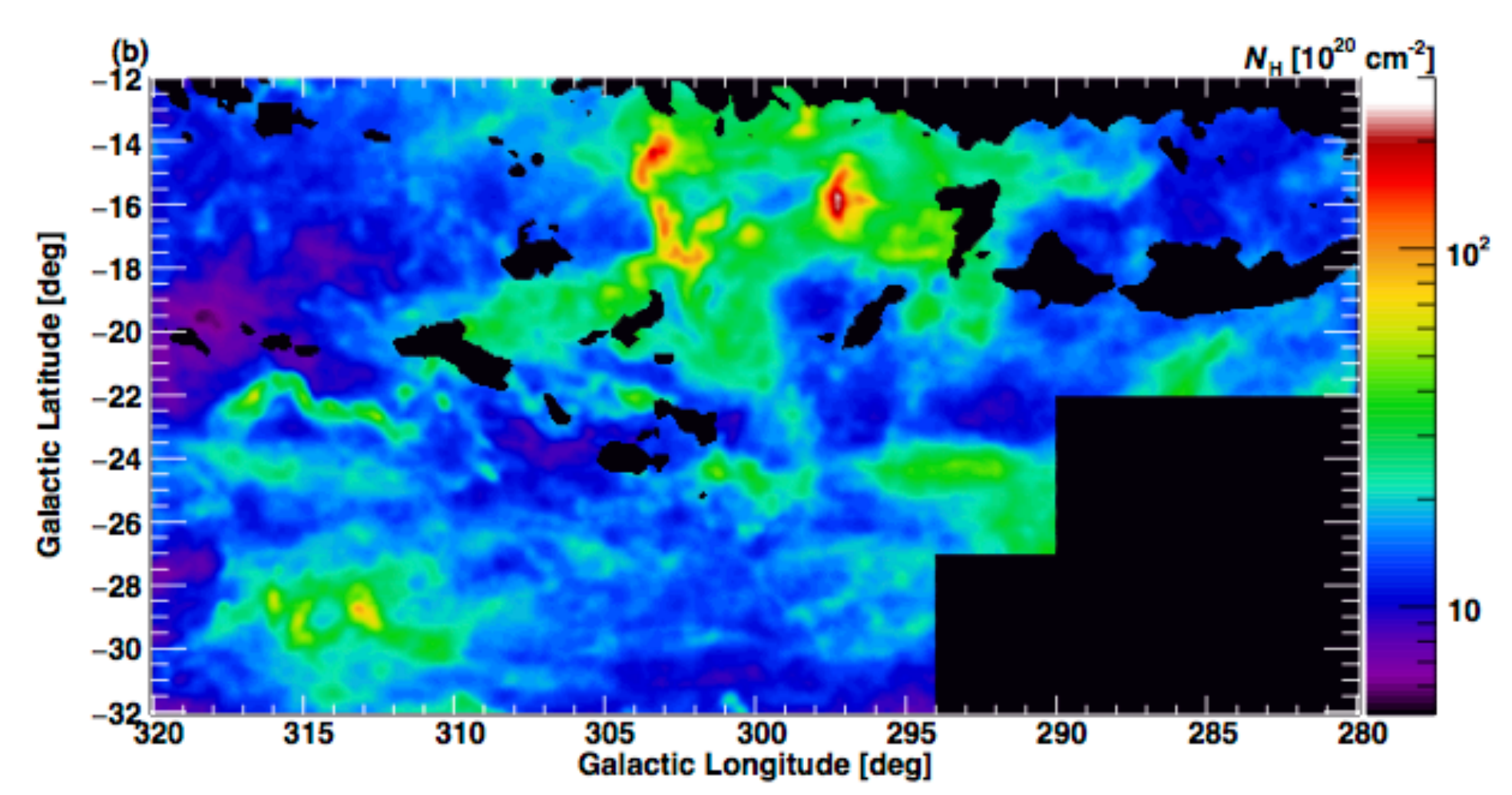}}}
   \end{center}
  \end{minipage} \\
  \end{tabular}  
   \caption{(a) Atomic ($\NHI$) and (b) total hydrogen ($\NH$) column density maps. The CO-emitting region is masked in the $\NHI$ map.}
\label{fig:fig9}   
\end{figure*}

Figures~\ref{fig:fig9}(a) and (b) respectively show distributions of $\NHI$ generated with the $\tauHI$ and $\Ts$ values and $\NH$ expressed by Equation (\ref{eq:NHmodelCham}). 
In the $\NHI$ map, the CO-dominated region is masked. 
The high column $\HI$ around the molecular clouds ($\NHI$ $\gtrsim$ 2$\times$10$^{21}$ cm$^{-2}$) indicates a large amount and extent of the optically thick $\HI$.
Its distribution is similar to those of high $\taud$ ($\gtrsim$ 1$\times$10$^{-5}$) and low $\Td$ ($\lesssim$ 18 K) (see Figure~\ref{fig:fig1}).
Physical association between the $\HI$ gas and dust properties are confirmed in the spatial distribution.

\begin{figure*}[h]
 \begin{tabular}{cc}
  \begin{minipage}{0.5\hsize}
   \begin{center}
    \rotatebox{0}{\resizebox{9cm}{!}{\includegraphics{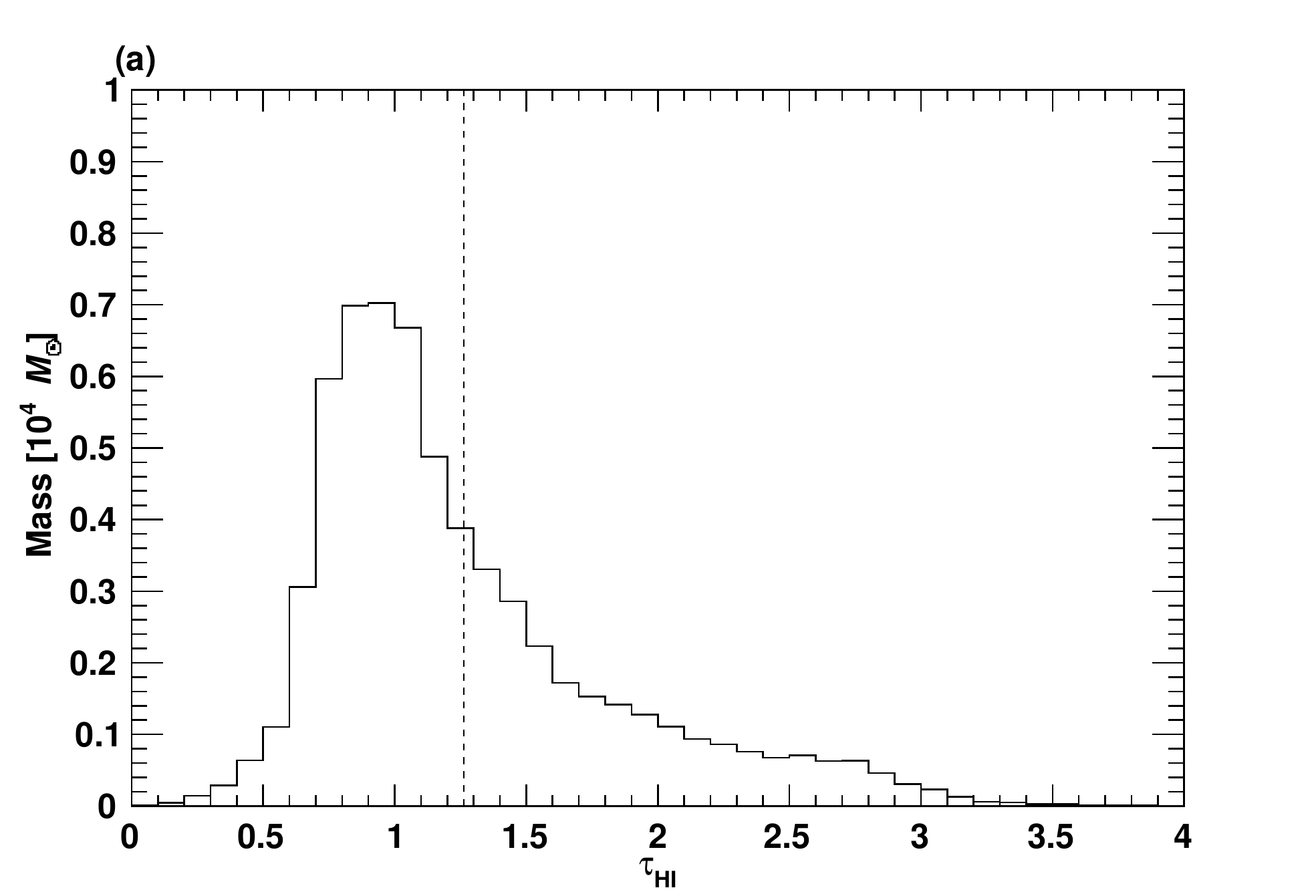}}}
   \end{center}
  \end{minipage} 
  \begin{minipage}{0.5\hsize}
   \begin{center}
    \rotatebox{0}{\resizebox{9cm}{!}{\includegraphics{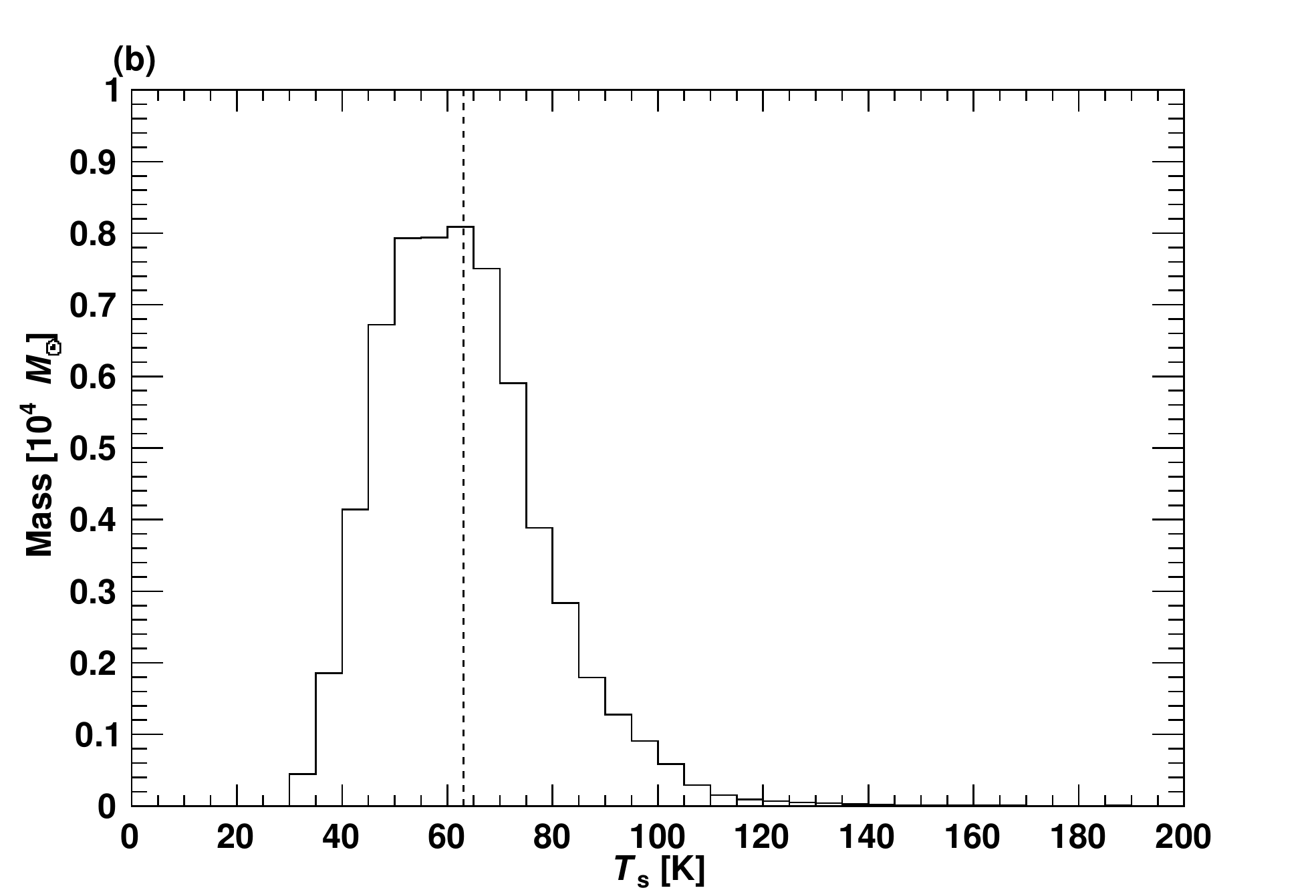}}}
   \end{center}
  \end{minipage} \\
  \begin{minipage}{0.5\hsize}
   \begin{center}
    \rotatebox{0}{\resizebox{9cm}{!}{\includegraphics{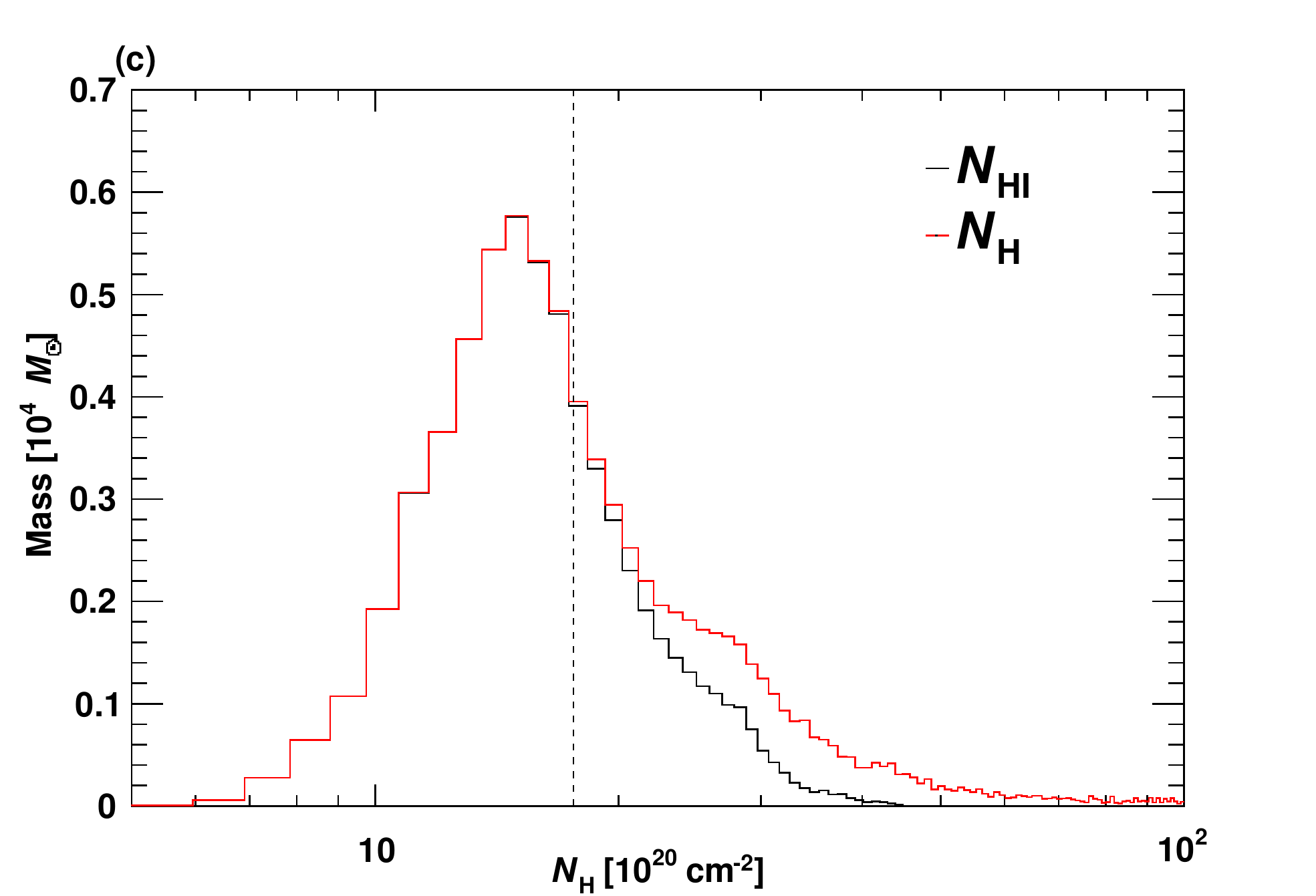}}}
   \end{center}
  \end{minipage} 
  \begin{minipage}{0.5\hsize}
  \begin{center}
   \end{center}
  \end{minipage} \\
  \end{tabular}  
   \caption{Mass-weighted histograms of (a) $\tauHI$, (b) $\Ts$ and (c) $\NHI$ (black) and $\NH$ (red). The vertical dashed lines indicate the mass-weighted mean values, $\langle\tauHI\rangle$ $=$ 1.3, $\langle\Ts\rangle$ $=$ 63 K and $\langle\NHI\rangle =$ 1.8 $\times$ 10$^{21}$ cm$^{-2}$.}
\label{fig:fig10}   
\end{figure*}

Figures~\ref{fig:fig10}(a), (b) and (c) show mass-weighted histograms of $\tauHI$, $\Ts$ and $\NHI$ in the analysis region, respectively.
The obtained $\NH$ histogram is also plotted in the panel (c).
For the mass calculation, we have assumed the distance for the Chamaeleon region 150 pc and adopted the mean atomic mass per H atom, $\mu =$ 1.41 \citep{Dappen00}.
Average values for $\tauHI$ and $\Ts$ are $\sim$1.3 and $\sim$63 K, respectively, which suggests large optical depth in the $\HI$-dominated region.
The $\NHI$ spans $\sim$(0.5--4) $\times$ 10$^{21}$ cm$^{-2}$ with the average value of $\sim$1.8 $\times$ 10$^{21}$ cm$^{-2}$.
The typical $\Htwo$ column density estimated from subtraction of $\NHI$ from $\NH$ is $\sim$1.5$\times$10$^{21}$~cm$^{-2}$, which is consistent with that obtained by the numerical simulation (Figure 4 in \citealt{Fukui+18}).
Gas masses of the atomic hydrogen excluding the CO-emitting region and the total hydrogen including the molecular gas are $\sim$6.3~$\times$~10$^{4}$ $\Msolar$ and $\sim$8.3~$\times$~10$^{4}$ $\Msolar$, respectively.
Our result indicates that the total mass is $\sim$30\% larger than the estimates in \citet{Planck15}.
The subtraction of the mass of the atomic gas component from the total mass, $\sim$2.0~$\times$~10$^{4}$ $\Msolar$, is much larger than the molecular gas mass $\sim$5000--8000 $\Msolar$ estimated by previous studies (e.g., \citealt{Mizuno+01}; \citealt{Ackermann+12}; \citealt{Planck15}).
This can be understood by the $\HI$ gas probably distributed in front and behind the molecular clouds.

\citet{Planck15} and \citet{Remy+18} pointed that the large gas column density inferred from the optically thick $\HI$ scenario should lower the local $\gamma$-ray emissivity, and thus yield a cosmic-ray density different from results of direct cosmic-ray measurements. 
However, our $\taud$-based $\NH$ model with the nonlinear effect lowered the column density compared to the linear relation with $\taud$ (F14; F15) and derived the total gas mass different from \citet{Planck15} by $\sim$30\%, which is within uncertainty of the measurements of the local $\gamma$-ray emissivity (see Table E.1 in \citealt{Planck15}).
This result indicates that our $\taud$-based $\NH$ model is acceptable in terms of the $\gamma$-ray studies.

Here we discuss an insight of the CO-dark $\Htwo$ in case of our $\taud$-based $\NH$ model.
If we assume that all the $\HI$ emission is optically thin, the mass of the atomic gas for the Chamaeleon region is derived to be $\sim$4.1$\times$10$^{4}$ $\Msolar$. 
Given that the CO-dark $\Htwo$ contributes to 50\% of the dark gas, the molecular gas mass for our derived total gas mass is estimated to be $\sim$25\% ($\sim$2$\times$10$^{4}$ $\Msolar$), which is 3--4 times larger than the gas mass traced by CO. 
This fraction is much larger than those estimated in other observational studies \citep{Planck15} and theoretical predictions (e.g., \citealt{Levrier+12}), where the CO-dark $\Htwo$ is a dominant component of the dark gas.
Although the CO-dark $\Htwo$ increases the total gas budget as suggested by many studies of molecular line surveys, we have not found evidence for the large mass of molecular gas as the dark-gas component.

Our results support the optically thick $\HI$ scenario, but
we found a contradiction with previous studies of the local $\HI$ gas properties.
 \citet{Stanimirovic+14} derived that a CNM fraction for Perseus molecular cloud is $\lesssim$~0.5 at the dark-gas medium where $\NHI$~$\gtrsim$~1~$\times$~10$^{20}$~cm$^{-2}$ (see Figure 8 in that paper). 
This CNM fraction is almost consistent with \citet{Fukui+18}.
Meanwhile, our $\NH$ model shows $\tauHI$ $\gtrsim$~1 at the dark-gas medium and thus suggests that the CNM is dominant there, which is not consistent with the slightly lower CNM fractions estimated by \citet{Stanimirovic+14} and \citet{Fukui+18}.
The cause of this discrepancy is not clear.
To reveal the difference among these studies, further evaluations of our gas model (e.g., variations of $\tauHI$ and $\Ts$ due to uncertainty of the $\NH$ model and $\Delta \VHI$) are required, as well as a number of $\HI$ emission-absorption measurements for the dark-gas medium of the Chamaeleon region.

\subsection{$\XCO$ Distribution} \label{sec: Xco}

The CO-to-$\Htwo$ conversion factor, $\XCO$, is usually estimated as $\sim$(1--2)$\times$10$^{20}$ cm$^{-2}$ K$^{-1}$ km$^{-1}$ s in the Milky Way disk (e.g., \citealt{Bolatto+13}). 
Using our $\taud$-based $\NH$ model, we can calculate $\XCO$ from a correlation with $\WCO$ as descried below.
The total column density $\NH$ is expressed as a sum of the number of protons in $\NHI$ and $\NHtwo$,

\begin{eqnarray}
\NH = \NHI + 2\NHtwo.
\label{eq:TotalNH} 
\end{eqnarray}

\noindent
By substituting $\XCO$ ($\equiv$ $\NHtwo$/$\WCO$) into Equation (\ref{eq:TotalNH}), the correlation between $\WCO$ and $\NH$ is expressed as,

\begin{eqnarray}
\NH = (2 \XCO)\times \WCO + \NHI = ({\rm slope}) \times \WCO + ({\rm intercept}).
\label{eq:NH_WCO} 
\end{eqnarray}
The half value of the slope gives an $\XCO$ factor. 

\begin{figure}[h]
 \begin{center}
  \includegraphics[width=100mm]{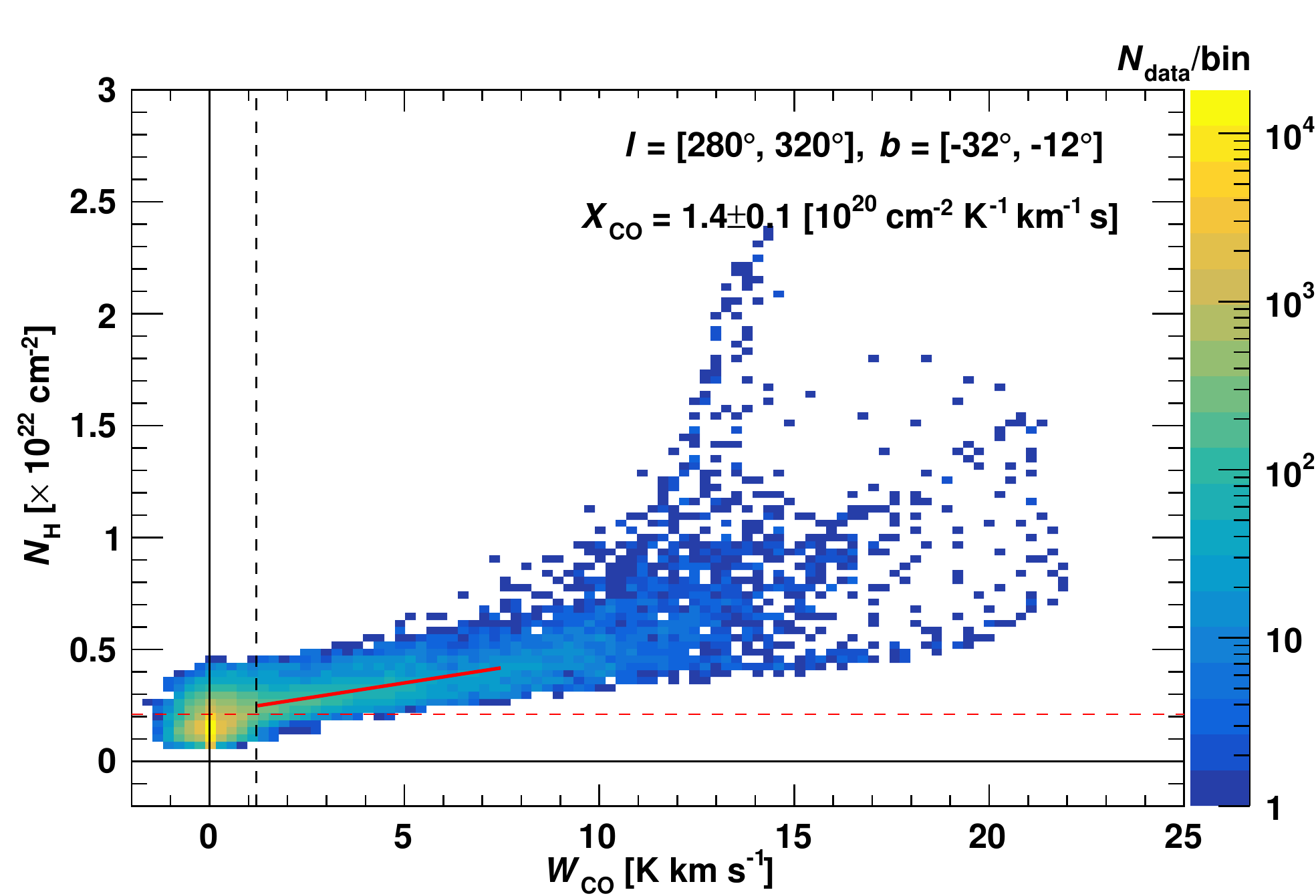}
  \end{center}
 \caption{$\WCO$--$\NH$ correlation plots for the all data points in the analysis region. The red solid line indicates the best-fit regression line obtained by the fit to the data points at 1.2~K~km~s$^{-1}$ (3 $\sigma$) $<$ $\WCO$ $<$ 7.5~K~km~s$^{-1}$. The vertical dashed line shows the lower limit of the fitting range. The horizontal red dashed line indicates the intercept value of $\NH$ obtained by the fitting.}
\label{fig:fig11}  
\end{figure}

\begin{figure*}[h]
 \begin{center}
  \includegraphics[width=175mm]{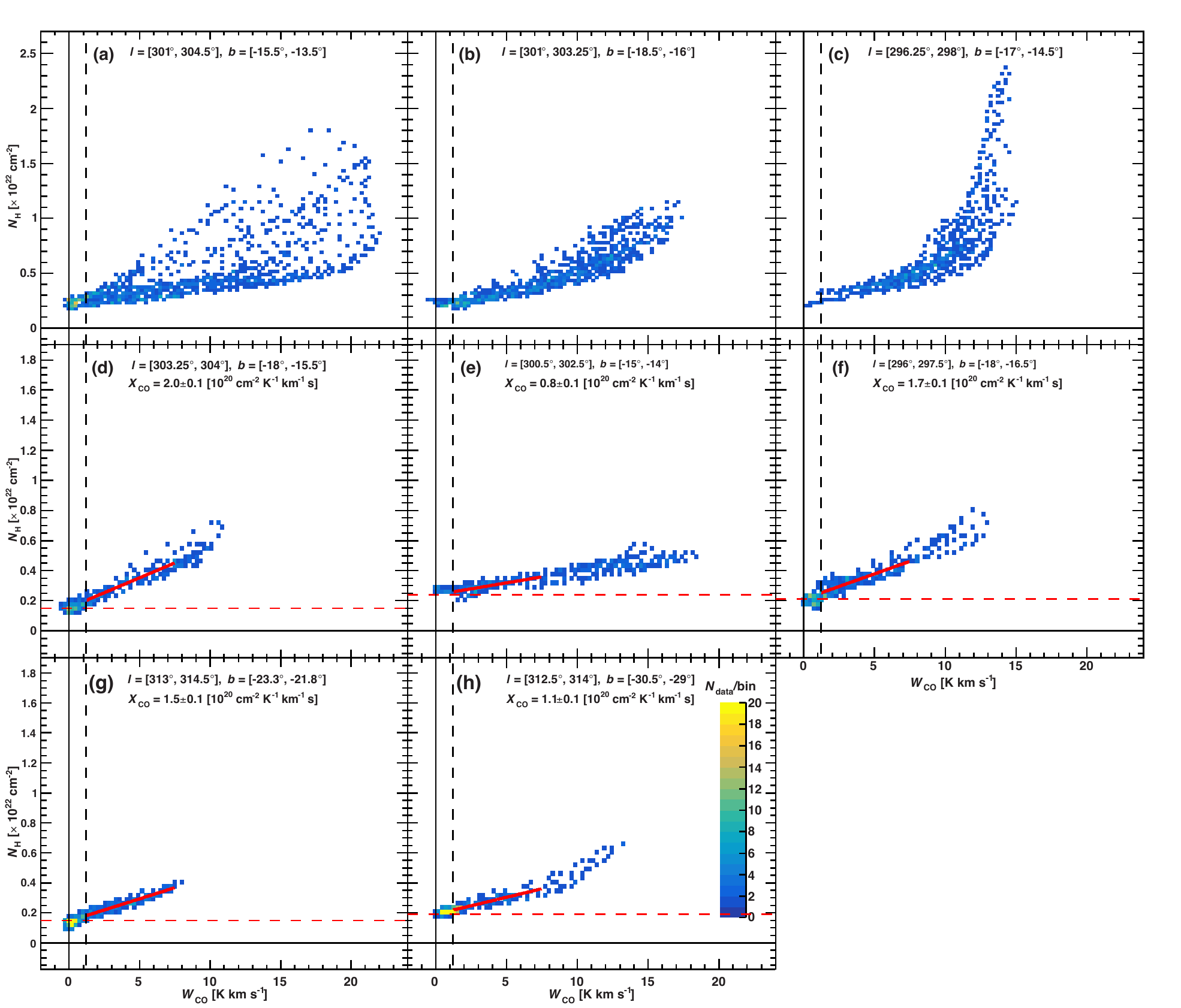}
  \end{center}
 \caption{$\WCO$--$\NH$ correlation plots with the data points taken from the areas (a)--(h) in Figure~\ref{fig:fig13}. The meaning of the lines in each panel are the same as Figure~\ref{fig:fig11}.}
\label{fig:fig12}  
\end{figure*}

\begin{figure*}[h]
 \begin{center}
  \includegraphics[width=150mm]{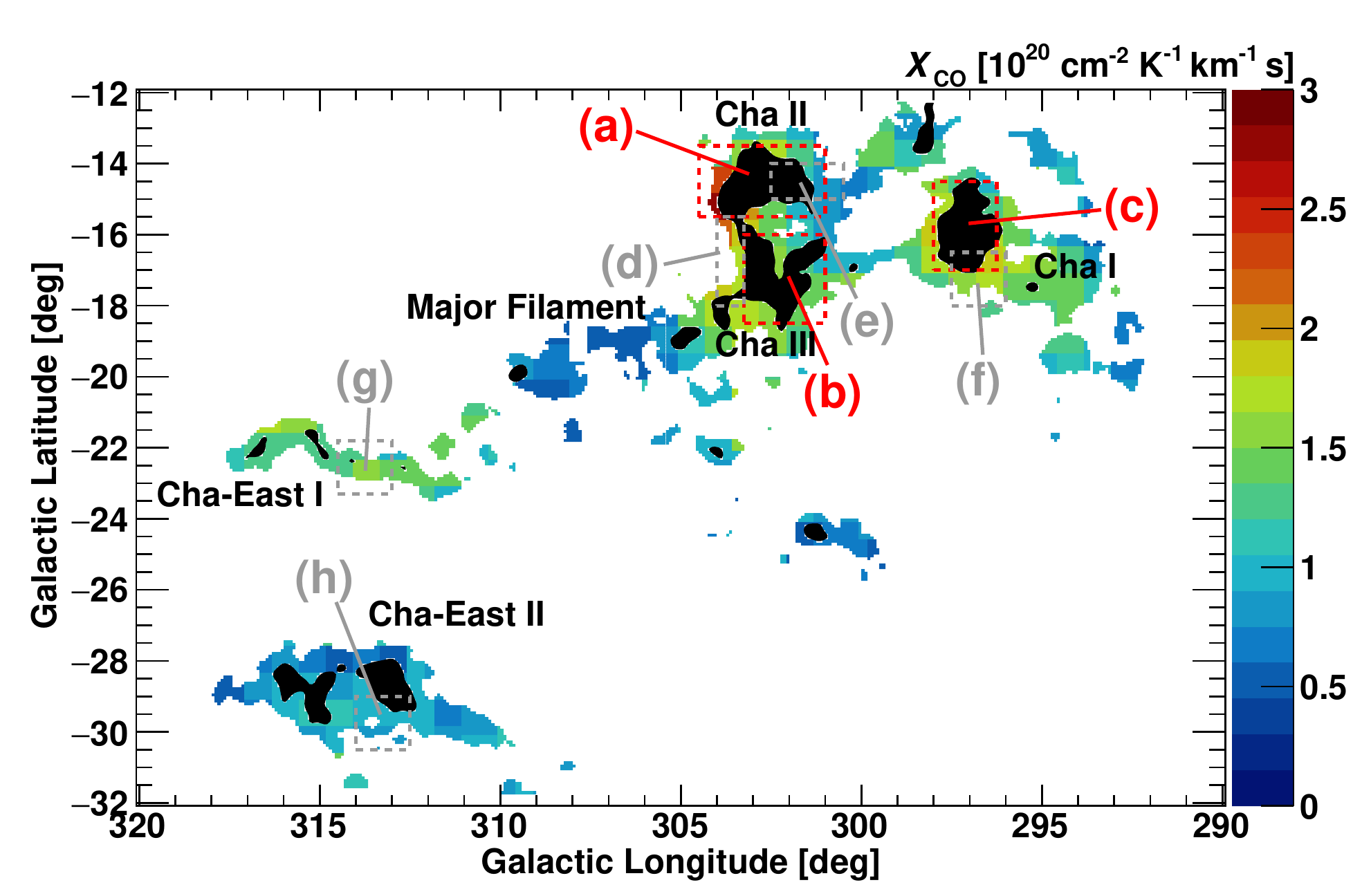}
  \end{center}
 \caption{$\XCO$ distribution derived from the correlation between $\WCO$ and $\NH$. 
 The areas (a)--(h) correspond to the correlation plots in Figures~\ref{fig:fig12}(a)--(h), respectively. The black-filled areas indicate regions with $\WCO >$~7.5~K~km~s$^{-1}$.
 The deriving method of the map is described in the text of Section~\ref{sec: Xco} and Figure~\ref{fig:fig20}.}
\label{fig:fig13}  
\end{figure*}

Figure~\ref{fig:fig11} shows a relationship between $\WCO$ and $\NH$ for the entire analysis region. 
One can see a positive correlation especially at low $\WCO$. 
However, significant deviation is seen at $\WCO \gtrsim$ 8~K~km~s$^{-1}$, probably due to saturation of the optically thick $^{12}$CO $J$$=$1--0 line often found in CO cores. 
In Figures~\ref{fig:fig12} (a)--(c), we make the same scatter plots whose data points are taken from the major clouds of the Chamaeleon region, Cha~I, II and III, which are covered by areas (a)--(c) as shown in Figure~\ref{fig:fig13}.
The significant deviation is found at high $\WCO$ especially in areas (a) and (c), which yield the large scattering 
in the $\WCO$-$\NH$ relationship in Figure~\ref{fig:fig11}.

Recent studies of the ISM suggest that the $\XCO$ have variations in a cloud complex depending on surrounding interstellar environment (e.g., \citealt{Bell+06}; \citealt{Lee+14}; \citealt{Okamoto+17}). 
To derive a spatial distribution of $\XCO$ from the $\WCO$-$\NH$ relationship, we divided the analysis region into 1$\fdg$5 $\times$ 1$\fdg$5 regions, and fit the data points with a linear function for each region that has the number of data points above 30 and the correlation coefficient above 0.7.
At first, the fitting range in $\WCO$ is restricted to from 1.2~K~km~s$^{-1}$ to 5~K~km~s$^{-1}$, to exclude contamination in $\WCO$ lower than the detection limit and the saturation effect at high $\WCO$.
Then we gradually raised the upper limit of the fitting range every 0.5~K~km~s$^{-1}$.
The values of $\XCO$ derived with $\WCO$ $<$ 7.5~K~km~s$^{-1}$ are consistent with those of $<$ 5~K~km~s$^{-1}$ within the statistical errors except for the regions where the data points at high $\WCO$ ($\gtrsim$ 8~K~km~s$^{-1}$) dominate.
We therefore adopted the $\XCO$ obtained from the fitting range at 1.2~K~km~s$^{-1}$ $<$ $\WCO$ $<$ 7.5~K~km~s$^{-1}$.
The derived 1$\fdg$5 $\times$ 1$\fdg$5-based $\XCO$ map is shown in the panel (a) of Figure~\ref{fig:fig20}, having some blanks in the diffuse medium where the value of $\XCO$ cannot be determined due to low statistics of the data and/or poor correlation of the $\WCO$-$\NH$ relationship.
To compensate the low data statistics, we extended the size of the fitting area to 2$^{\circ}$$\times$2$^{\circ}$ and 2${\fdg}$5$\times$2${\fdg}$5 regions, whose results are shown in the panels (b) and (c) of Figure~\ref{fig:fig20}, respectively.
In these maps, $\XCO$ for the blank regions found in the 1$\fdg$5 $\times$ 1$\fdg$5-based map is given owing to increase of the data points.
On the other hand, the larger pixelized map possibly overlooks local variation of $\XCO$ such as a region at ($l$, $b$) $\sim$ (304$^{\circ}$, $-$14$\fdg$5), where the $\XCO$ values of the 1$\fdg$5 $\times$ 1$\fdg$5-based map are significantly different from the other two maps.
We therefore combined the three $\XCO$ maps, preferentially adopting the $\XCO$ from the 1$\fdg$5 $\times$ 1$\fdg$5-based map; 
the other two maps are used to compensate the blanck regions. 
The finally obtained $\XCO$ map is shown in Figure~\ref{fig:fig13}. 
The $\XCO$ spans $\sim$(0.5--3) $\times$ 10$^{20}$ cm$^{-2}$~K$^{-1}$~km$^{-1}$~s across the cloud complex and the average one derived from Figure~\ref{fig:fig11} is 1.4 $\pm$ 0.1 $\times$ 10$^{20}$ cm$^{-2}$~K$^{-1}$~km$^{-1}$~s. 
These values are consistent with the typical $\XCO$ measured for the Galactic interstellar clouds, (1--2) $\times$ 10$^{20}$cm$^{-2}$~K$^{-1}$~km$^{-1}$~s (e.g., \citealt{Bolatto+13}). 
A study of the ISM of the Chamaeleon region, \citet{Planck15}, also derives an $\XCO$ factor comparable to our result through their analysis using the dust optical depth, $\taud$.  

In Figures~\ref{fig:fig12} (d)--(h), we show the correlation plot for the regions enclosed by the dashed lines in the $\XCO$ map in Figure~\ref{fig:fig13}.
These regions are selected to more clearly show and discuss possible relations between the $\XCO$ and the surrounding gas condition. 
We fit the data points with a linear function similarly to Figure~\ref{fig:fig11} and found that each region shows a good correlation, but their slopes vary depending on their positions.
In addition to the $\XCO$, the fitting also gives the intercept values of $\NHI$, as represented by the red-dashed lines in Figure~\ref{fig:fig12} (also in Figure~\ref{fig:fig11} as an average value for the entire analysis region). 
These values correspond to average $\HI$ column density in each fitted region and are comparable to the $\NHI$ around the masked CO-dominated region shown in Figure~\ref{fig:fig9}(a).

The Chamaeleon complex consists of Cha~I--III, Cha-East~I, Cha-East~II, and Major Filament, having different evolutionary history and star formation activity \citep{Mizuno+01}. 
We found $\XCO$ at 303$^{\circ}$~$\lesssim l \lesssim$~304$^{\circ}$ in Cha~II and Cha~III, and the whole cloud of Cha~I show relatively high values, $\sim$(1.5--3) $\times$ 10$^{20}$cm$^{-2}$~K$^{-1}$~km$^{-1}$~s, as shown by the correlations for areas (d) and (f). 
Cha-East~I including area (g) also has a relatively high $\XCO$ ($\sim$1.5 $\times$ 10$^{20}$ cm$^{-2}$~K$^{-1}$~km$^{-1}$~s.). 
Conversely, $\XCO$ at 300$^{\circ}$~$\lesssim l \lesssim$~301$^{\circ}$ in Cha~II, where a part of area (e) is included, is relatively small ($\lesssim$ 1 $\times$ 10$^{20}$ cm$^{-2}$~K$^{-1}$~km$^{-1}$~s). 
Cha-East~II and Major Filament also show similar low $\XCO$. 
The $\NHI$ map in Figure~\ref{fig:fig9}(a) indicates that areas (d) and (g) are faced to the interstellar space with relatively low column density ($\sim$1.5 $\times$ 10$^{21}$ cm$^{-2}$), while the $\NHI$ surrounding area (e) and Major Filament is rather high ($\sim$2--3.5 $\times$ 10$^{21}$ cm$^{-2}$).
The large $\XCO$ in areas (d) and (g) is probably due to CO destruction in the low-density medium, while the small $\XCO$ in area (e) could be ascribed by sufficient dust shielding with the visual extinction, $A_{\rm v}$ $\gtrsim$~1 mag ($\AJ$ $\gtrsim$~0.3 mag), preventing CO from photodissociation.
The tend of high $\XCO$ in regions with low CO abundance outside molecular clouds is consistent with observational studies of molecular cloud regions (e.g., \citealt{CottenMagnani13}; \citealt{Okamoto+17}) and theoretical predictions of formation of molecular clouds (e.g., \citet{InoueInutsuka12}). 

Among the six cloudlets in the Chamaeleon complex, Cha~I exhibits the highest star formation activity represented by two Ae/Be stars located in the CO core \citep{Luhman08}. 
The relatively high $\XCO$ around the CO cores in Cha~I, exemplified by area (f) ($\XCO$ $\sim$1.7 $\times$ 10$^{20}$cm$^{-2}$~K$^{-1}$~km$^{-1}$~s), may be ascribed to the more intense radiation field and enhanced stellar feedback, suggesting preferential destruction of CO in the star-forming region. 
Cha~II holds a few tens of young stellar objects at 303$^{\circ}$~$\lesssim l \lesssim$~304$^{\circ}$ and $-$15$^{\circ}$~$\lesssim b \lesssim$~$-$13$\fdg$5 \citep{Alcala+08}, which nearly corresponds to the positions with the highest $\XCO$ $\sim$3.0 $\times$ 10$^{20}$cm$^{-2}$~K$^{-1}$~km$^{-1}$~s, possibly showing the gas properties related to an evolutionary step of the low-mass stars. 
On the other hand, Cha-East~II has relatively lower $\XCO$ as seen in the correlation for area (h). 
This result makes sense in terms of no strong radiation field in the surrounding medium and lack of star-forming activity in the CO cores of the clouds.

\begin{figure*}[h]
 \begin{center}
  \includegraphics[width=100mm]{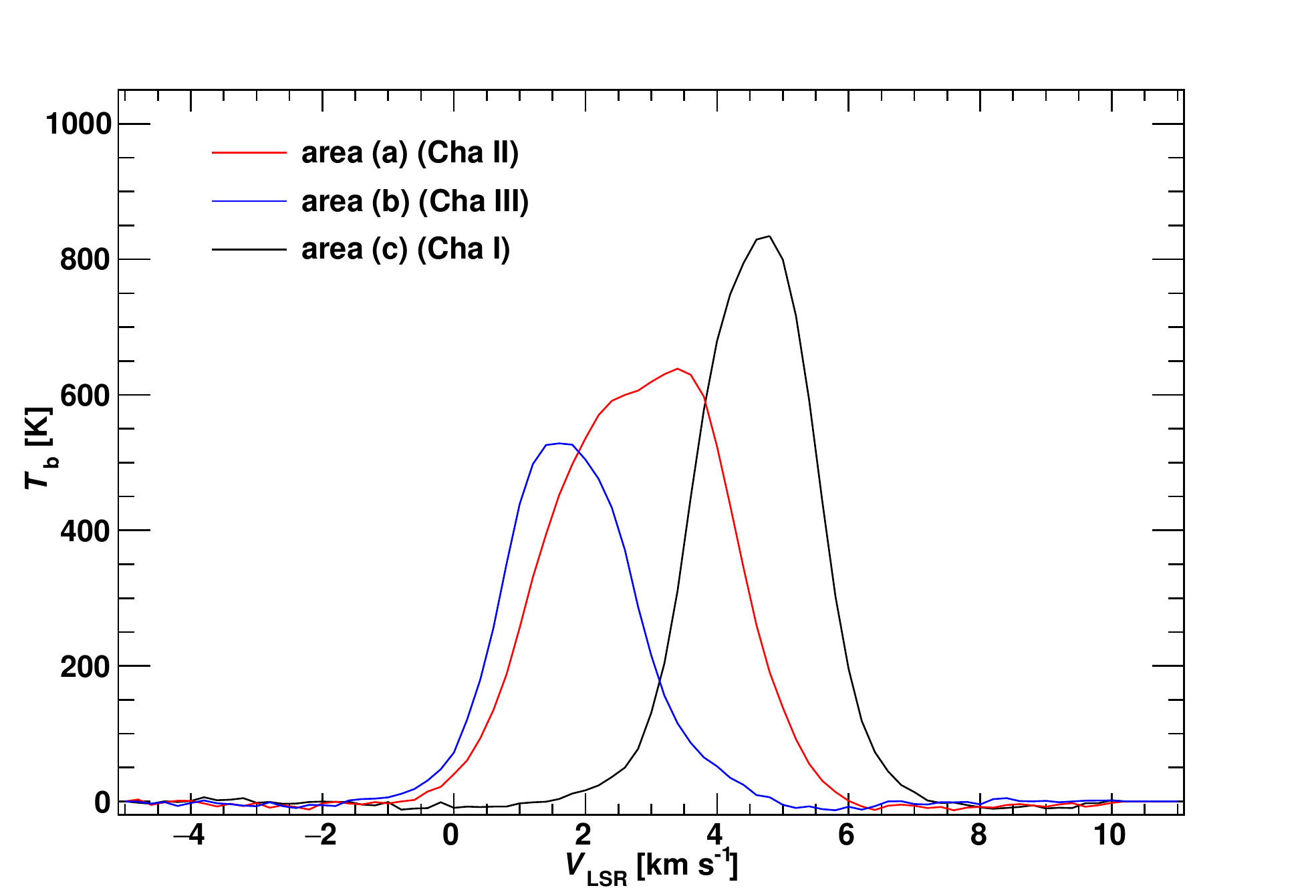}
  \end{center}
 \caption{The spectra of $^{12}$CO $J =$1--0 for areas (a), (b) and (c) shown in Figure~\ref{fig:fig13}. The brightness temperature ($T_{\rm b}$) on the y-axis is a sum of $T_{\rm b}$ of each pixel of the $\WCO$ map in Figure~\ref{fig:fig1} (a).}
\label{fig:fig13p5}  
\end{figure*}

Finally, we note a peculiar correlation found in the CO core of Cha II. We found that the $\WCO$--$\NH$ relationship in area (e) shows a tight correlation without the $\WCO$ saturation, in spite of inclusion of the data points at high $\WCO$ ($\gtrsim$ 10~K~km~s$^{-1}$).
This trend is also confirmed in the $\taud$--$\WCO$ relationship of Figures~\ref{fig:fig5}(b) and \ref{fig:fig18}, as a scatter distribution at $\taud$ $\sim$ 4$\times$10$^{-5}$ and $\WCO$ $>$ 10~K~km~s$^{-1}$.
One of the possibility of the high value of $\WCO$ is due to a high CO abundance in this region, suggesting possible different age of the molecular gas and a different evolutionary history of the clouds in the Chamaeleon region.
The other possibility of the lower saturation is opacity effect of the CO line.
Figure~\ref{fig:fig13p5} shows CO spectra for the Cha I-III regions, whose intensities are summed for the pixels in areas (a)--(c). 
We confirmed that the spectrum of the Cha II (area (a)) is more broaden than those of the others, which may suggest gas turbulence leading to lower the opacity of the cloud. 
A few tens of young stellar objects are included in the Cha II region \citep{Alcala+08} and thus possible outflows may contribute this effect, as similarly suggested in a star-forming region of the Perseus molecular cloud \citep{Pineda+08}.

\clearpage

\subsection{Comparison with Other Molecular Cloud Regions}

\begin{figure*}[h]
 \begin{center}
  \includegraphics[width=120mm]{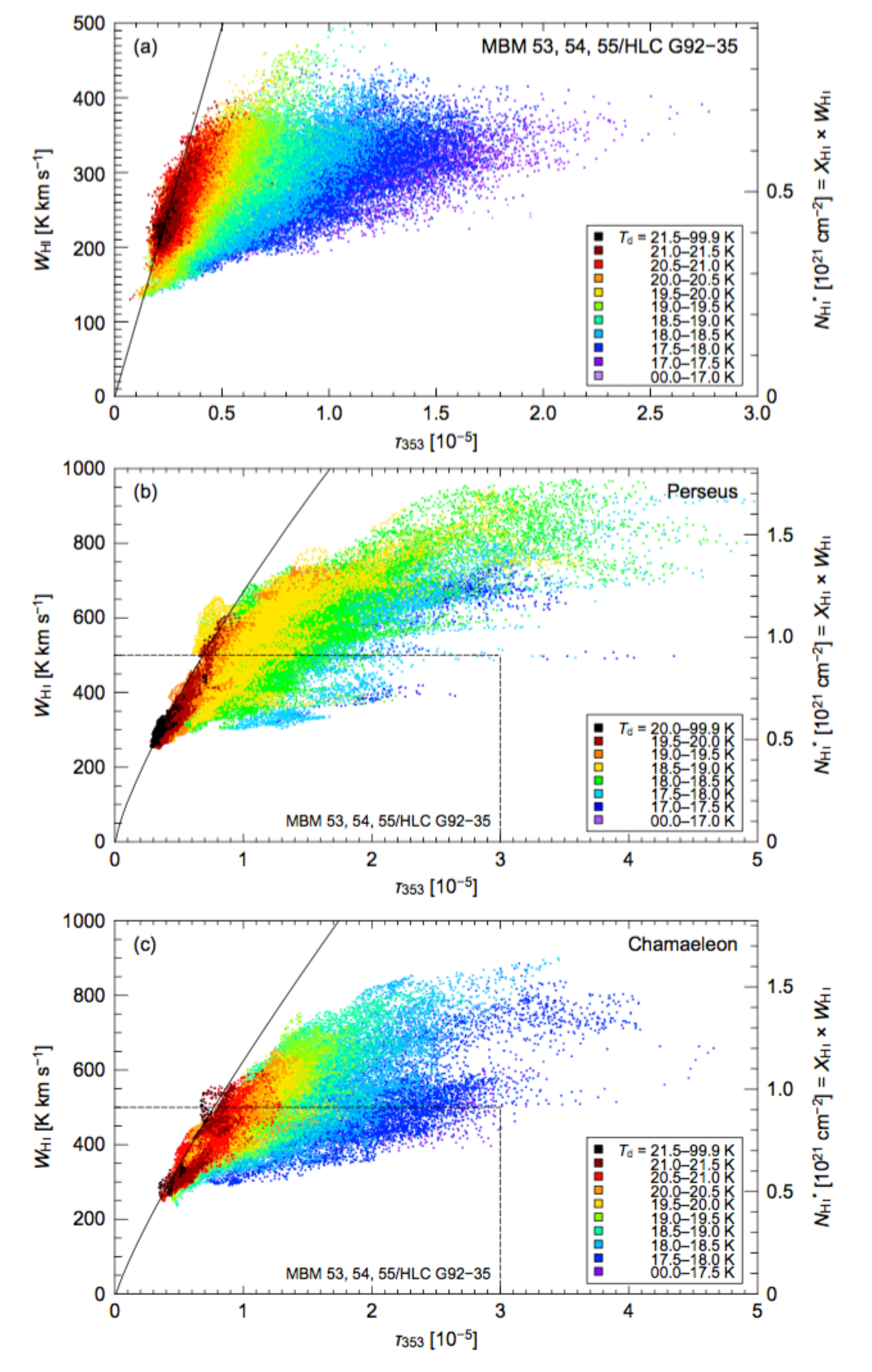}
  \end{center}
 \caption{$\taud$--$\WHI$ correlation plots for the (a) MBM~53--55, (b) Perseus, and (c) Chamaeleon regions. The black solid line/curves indicate the $\WHI$ theoretical model as a function of $\taud$ (Equation (\ref{eq:WHIModelCurve})) in the case of $\tauHI \ll$~1, $\tauHI =$~0.21 (see Appendix A in \citealt{Okamoto+17}), and 0.34 (see Appendix \ref{sec:RefPoints}) for the MBM~53--55, Perseus and Chamaeleon regions, respectively. For easier comparison, the horizontal and vertical ranges in the panel (a) are shown in the panels (b) and (c).}
\label{fig:fig14}  
\end{figure*}

Finally, we compare gas properties among local molecular cloud complexes, MBM~53--55 (F14), Perseus \citep{Okamoto+17}, and Chamaeleon (this study) regions, for which dedicated analyses have been performed with our $\taud$-based $\NH$ model.  
The $\taud$--$\WHI$ correlation plots of each region are shown in Figures~\ref{fig:fig14}(a)--(c).
For easier comparison, the horizontal and vertical ranges in the panel (a) are shown in the panels (b) and (c).
Table~\ref{tab:Parameters_of_regions} summarizes physical quantities obtained by these studies. 

$\taud$ in MBM~53--55 region is approximately one order of magnitude smaller than those in the other two regions, which yields relatively lower $\NH$. 
The large $\taud$ in the Perseus and Chamaeleon regions gives larger $\NH$, although it is unlikely to increase with a simple linear function of $\taud$.
We have found a nonlinear relation with $\alpha \sim $1.2--1.3 in the $\taud$--$\AJ$ relationship for the dense molecular clouds ($\AJ \gtrsim$ 0.3 mag) of the Perseus and Chamaeleon regions. 
This nonlinear relation also traces the mild curvature in the $\taud$--$\WHI$ relationship for the $\HI$-dominated medium, which is clearly different from the linear relation seen in the MBM~53--55 region (see Figure~\ref{fig:fig14}).
These variations of dust opacity may arise from different grain evolution among the cloud complexes.
Taking into account the dust evolution effect, the $\NH$ for the Perseus and Chamaelelon regions are found to be larger by factor of $\sim$2--6 than that for the MBM~53--55 region.

The total column density model as a function of $\taud$ allows us to investigate the atomic and molecular gas properties.
The obtained $\langle\Ts\rangle$ is the highest in Perseus region and it becomes lower followed by Chamaeleon and MBM~53--55 regions.
The trend of dust evolution ($\alpha$) and the $\HI$ gas properties ($\Ts$ and $\tauHI$) might be associated with current star-forming activities: less star formation in the MBM~53--55 clouds (e.g., \citealt{Yamamoto+03}), high-mass star-formation in the Perseus (e.g., \citealt{Bally+08}) and low-mass star-formation in the Chamaeleon (e.g., \citealt{Luhman08}) regions.
The Perseus molecular clouds are included in the Perseus OB2 Association, which forms a part of Gold’s Belt.
Massive stars located in this region may heat the surroundings atomic gas and generate relatively higher $\Ts$.
The MBM~53--55 region shows the largest $\langle\tauHI\rangle$, which indicates a large amount of dark gas between the $\HI$ and CO transition.
This is consistent with the large mass fraction of dark gas in this region suggested by a $\gamma$-ray analysis \citep{Mizuno+16}.
The relatively lower $\alpha$ suggests less dust evolution, which may relate to the lower current star-formation activity. 
The $\XCO$ becomes smaller in the MBM~53--55 region and larger in the Perseus region.
According to recent studies of $\XCO$ in the local interstellar clouds, $\XCO$ in high-density regions tends to be lower (e.g., \citealt{CottenMagnani13}; \citealt{Schultheis+14}).
A comparison with the $\WCO$ (peak) follows this tendency.

Among the three regions, the Chamaeleon region exhibits an intermediate $\HI$ and CO gas properties between the MBM~53--55 and Perseus regions.
The lack of OB clusters in the Chamaeleon region yields relatively quiet environments, whereas low-mass star formations found in the CO cores may affect gas properties in the opaque regions.

\begin{table*}[h]
\caption{Physical quantities to represent gas properties of the MBM~53--55, Perseus, and Chamaeleon regions.} 
\label{tab:Parameters_of_regions}
 \begin{center}
\begin{tabular}{cccccccc} \hline\hline
Region & $\taud$ & $\alpha$ & $\NH$ & $\langle\tauHI\rangle$ & $\langle\Ts\rangle$ & $\WCO$(peak) & $\langle\XCO\rangle$\\ 
 &  &  & $[\UCND]$ &  & $[\UK]$ & [K km s$^{-1}$]  & $[\UXCO]$\\ \hline
MBM~53--55 & ($7\text{--}600)\times10^{-7}$ & $1.0$ & ($1\text{--}100)\times10^{20}$ & $1.5$ & $46$ & 21 & $1.8\times10^{20}$\\ 
Perseus & ($3\text{--}800)\times10^{-6}$ & $1.3$ & ($5\text{--}400)\times10^{20}$ & $0.9$ & $97$ & 70 & $1.0\times10^{20}$\\ 
Chamaeleon & ($3\text{--}300)\times10^{-6}$ & $1.2$ & ($6\text{--}200)\times10^{20}$ & $1.3$ & $62$ & 25 & $1.4\times10^{20}$\\ \hline
\multicolumn{8}{l}{The brackets indicate the average value for the parameters.}\\
\end{tabular}
\end{center}
\end{table*}

\clearpage

\section{Conclusions} \label{sec:conclusion}

As part of an analysis of interstellar hydrogen gas based on the {\it Planck} data we carried out a comparative study of $\HI$, CO and dust in the Chamaeleon molecular cloud complex. 
The main conclusions are summarized below.

\begin{enumerate}
\item A comparison of the $J$-band extinction $\AJ$ and submillimeter dust optical depth $\taud$ shows a relationship that $\taud$ increases as the $\sim$1.2th power of $\AJ$.
This indicates that the total column density $\NH$ is modeled by the $\sim$1/1.2th power of $\taud$, and suggests dust growth in dense molecular clouds.
Similar trends are found in the Perseus and Orion A clouds, whereas the index may vary within about $\pm$10\% from region to region.
\item We have found a scatter correlation between $\taud$ and $\WHI$, similar to those found in the MBM 53--55 (F14) and Perseus \citep{Okamoto+17} molecular cloud regions.
Applying the nonlinear relation found in the $\taud$--$\AJ$ relationship to the $\taud$-based $\NH$ model reproduces the scatter correlation, which indicates a large amount of the optically thick $\HI$ around the molecular clouds.
The average $\tauHI$ and $\Ts$ in the Chamaeleon region are derived to be $\sim$1.3 and $\sim$63 K, respectively.
\item A distribution of an $\XCO$ factor in the Chamaeleon complex was derived. 
We have found variation of $\XCO$ $\sim$(0.5--3) $\times$ $10^{20}$ cm$^{-2}$ K$^{-1}$ km$^{-1}$ s, which is consistent with the typical value in the Galaxy. 
This is possibly due to different physical conditions related to the surrounding ISRF.
\item Gas properties in the Chamaeleon region are compared with the MBM~53--55 and Perseus molecular cloud regions.
The Chamaeleon region has moderate $\tauHI$, $\Ts$ and $\XCO$ among the three regions.
The moderate ISRF in the diffuse medium and low-mass star-formation activities in the cores of clouds may relate to these gas properties.

\end{enumerate}

\acknowledgments

This work was financially supported by Japan Society for the Promotion of Science (JSPS) KAKENHI, grant numbers 15H05694, 25287035.
We acknowledge the use of the Legacy Archive for Microwave Background Data Analysis (LAMBDA), part of the High Energy Astrophysics Science Archive Center (HEASARC). 
HEASARC/LAMBDA is a service of the Astrophysics Science Division at the NASA Goddard Space Flight Center.
The {\it Planck} satellite, an ESA science mission, is supported by ESA Member States, USA, and Canada.
The CO data were obtained with the NANTEN telescope operated with a mutual agreement between Nagoya University and the Carnegie Institution of Washington. 
We also utilized all-sky $\HI$ map from the Parkes Galactic All-Sky Survey (GASS) and $\Halpha$ map \citep{Finkbeiner03}, combined with the data from Wisconsin H-Alpha Mappe (WHAM), Virginia Tech Spectral-Line Survey (VTSS) and Southern H-Alpha Sky Survey Atlas (SHASSA).
The $J$-band infrared extinction map based on the NICEST method \citep{JuvelaMontillaud16} and the 21 cm radio continuum map from CHIPASS \citep{Calabretta+14} were also utilized in this study.  
Some of the results in this paper have been derived using the HEALPix \citep{Gorski+05} package.

\software{HEALPix (Gorski et al. 2005)}

\clearpage
\appendix

\section{$\HI$ Gas Distribution Separated in Velocity} \label{sec: HIgasVelChannel}

\begin{figure*}[h]
 \begin{center}
  \includegraphics[width=150mm]{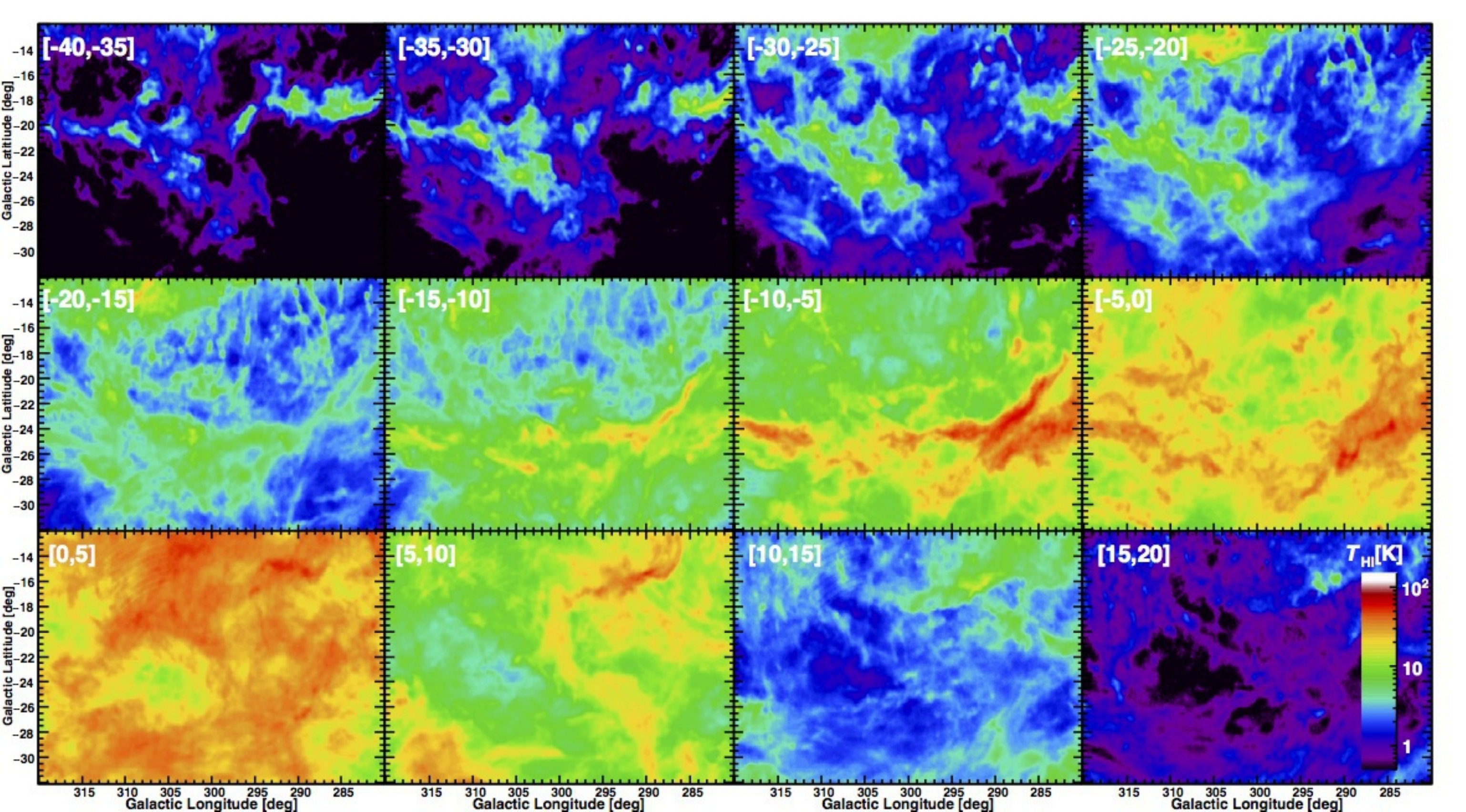}
  \end{center}
 \caption{$\HI$ velocity channel distribution from $-40$~km~s$^{-1}$ to $+20$~km~s$^{-1}$, whose intensities are averaged within the velocity intervals of 5~km~s$^{-1}$, giving the brightness intensity ($\THI$) in units of K.}
\label{fig:fig15}  
\end{figure*}

\begin{figure}[h]
 \begin{center}
  \includegraphics[width=100mm]{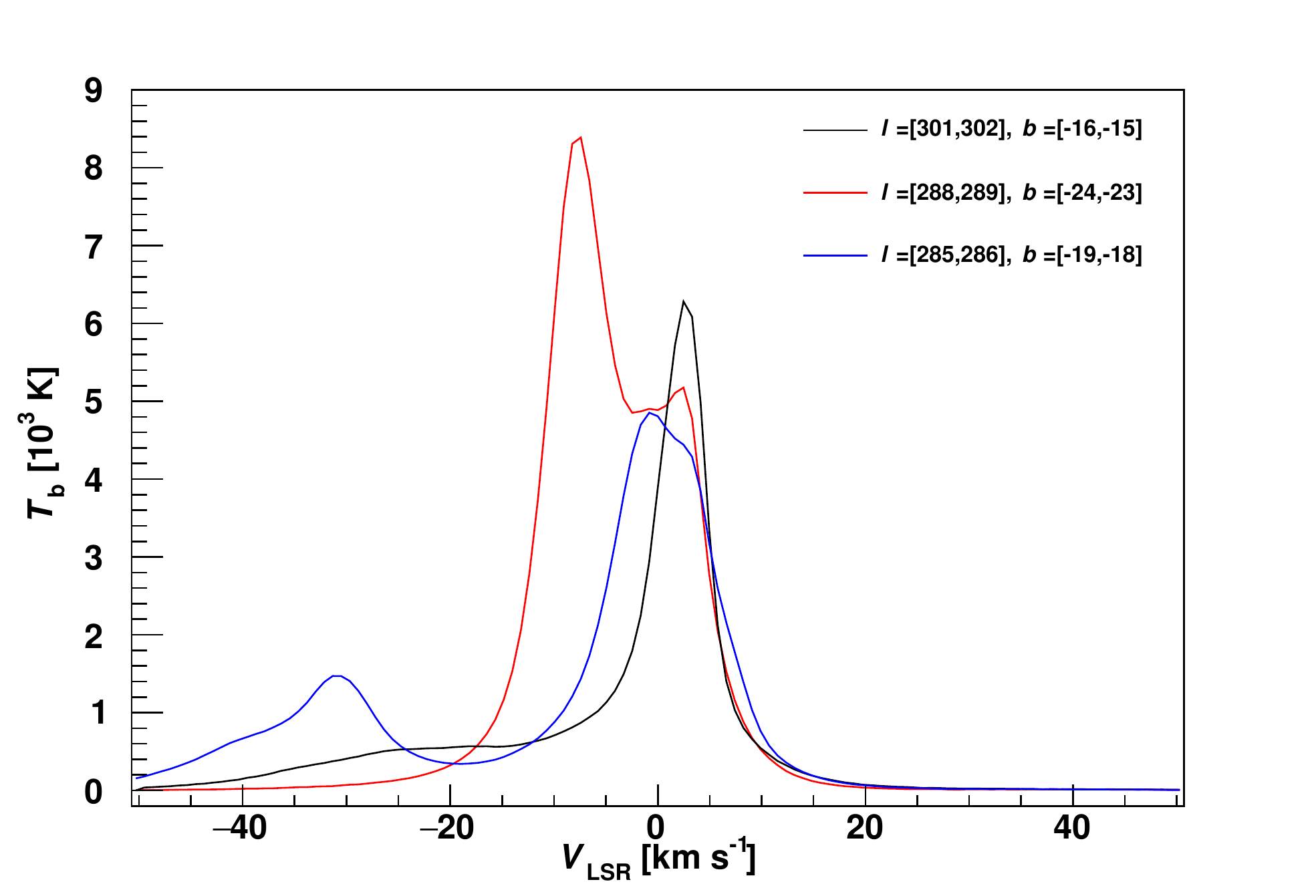}
  \end{center}
 \caption{Examples of the $\HI$ line profiles for the three regions with size of 1$\fdg$0 $\times$ 1$\fdg$0, having the characteristic line profiles of the local (black), IVA (red) and the high velocity (blue) components. The brightness temperature ($T_{\rm b}$) on the y-axis is a sum of $T_{\rm b}$ of each pixel. The regions with the red and blue spectra are included in the masked area.}
\label{fig:fig16}  
\end{figure}

Figure~\ref{fig:fig15} indicates $\HI$ velocity channel map from $-40$~km~s$^{-1}$ to $+$20~km~s$^{-1}$ separated into 5~km~s$^{-1}$ intervals, whose integrated intensities are averaged by the velocity range, which gives the $\HI$ brightness temperature ($\THI$).
These gas distributions exhibit roughly three structures,
\begin{itemize}
\item
local clouds extensively distributed at $-5$~km~s$^{-1}$ $\lesssim$ $\VLSR$ $\lesssim$ $+10$~km~s$^{-1}$
\item
elongated gas structure, crossing the whole region around $b \sim$~$-25^{\circ}$ at $-20$~km~s$^{-1}$ $\lesssim$ $\VLSR$ $\lesssim$ $-5$~km~s$^{-1}$ (IVA, see \citealt{Planck15})
\item
high velocity component located at 302$^{\circ}$~$\lesssim l \lesssim$~314$^{\circ}$ and $-30^{\circ}$~$\lesssim b \lesssim$~$-12^{\circ}$ seen at $\VLSR \lesssim -20$~km~s$^{-1}$.
\end{itemize}
In Figure~\ref{fig:fig16}, we give examples of the spectra showing these line profiles.
The red and blue spectra have strong emissions from the IVA and the high velocity components.
To avoid contamination from other than the local clouds, the IVA-dominated region at $l <$ 290$^{\circ}$ and $b <$ $-$22$^{\circ}$ and regions with the high velocity components are masked in the present study.
Strong emission from the LMC outskirts detected at $+200$~km~s$^{-1}$ $\lesssim \VLSR \lesssim$ $+300$~km~s$^{-1}$ is also dropped by the mask (c) (see Figure~\ref{fig:fig2}). 

\clearpage

\section{Correlations of $\WHI$ and $\WCO$ with $\taud$} \label{sec: CorrWHIWCOandtau353}

\begin{figure*}[h]
 \begin{center}
  \includegraphics[width=180mm]{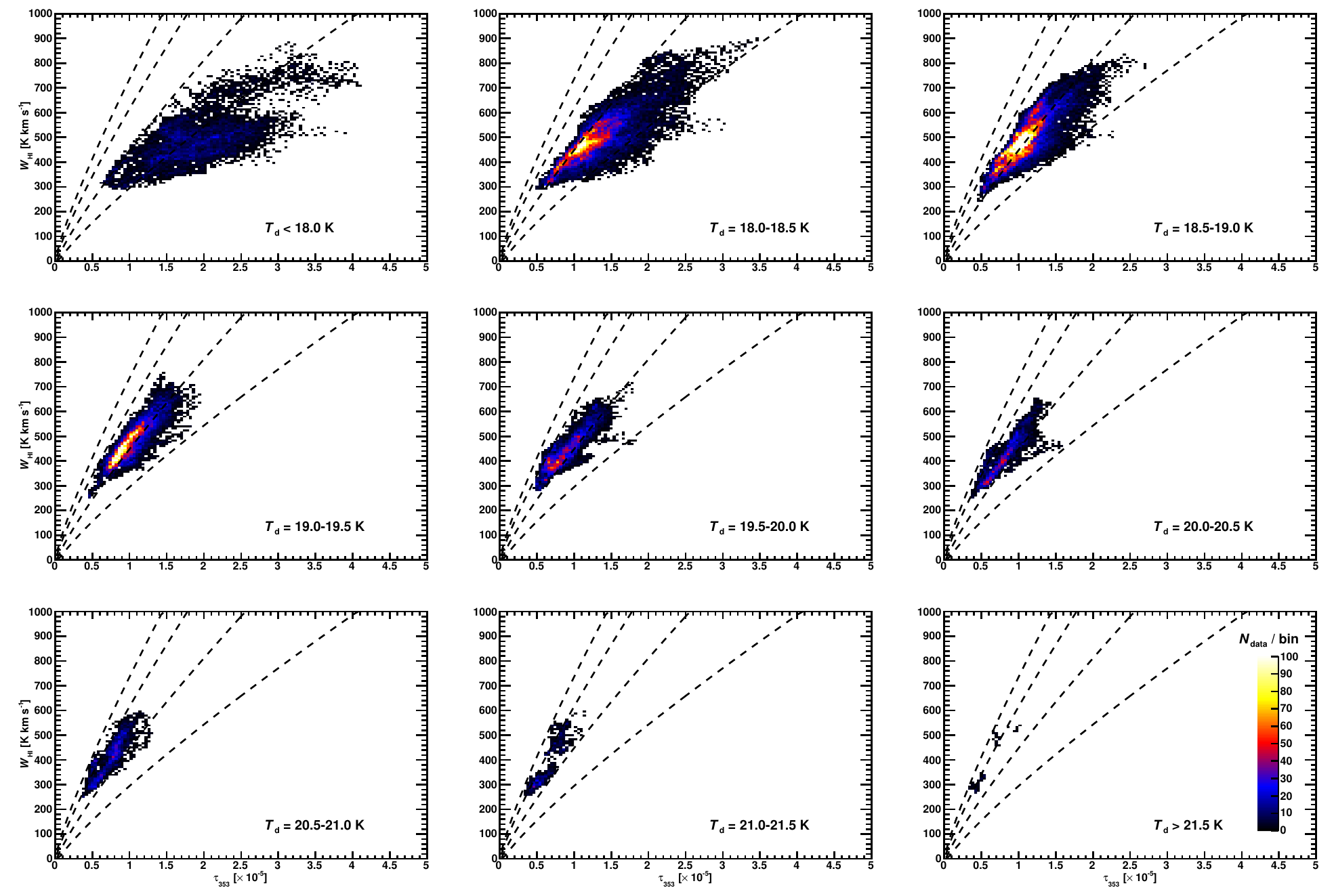}
  \end{center}
 \caption{Correlation plots (in density) between $\taud$ and $\WHI$ sorted by several $\Td$ intervals. The dashed curves show the theoretical functions of Equation~(\ref{eq:WHIModelCurve}) for $\alpha =$~1.2, with $\tauHI \ll$ 1, $\tauHI =$ 0.34, 1.0 and 2.0 from left to right.}
\label{fig:fig17}  
\end{figure*}

\begin{figure*}[h]
 \begin{center}
  \includegraphics[width=180mm]{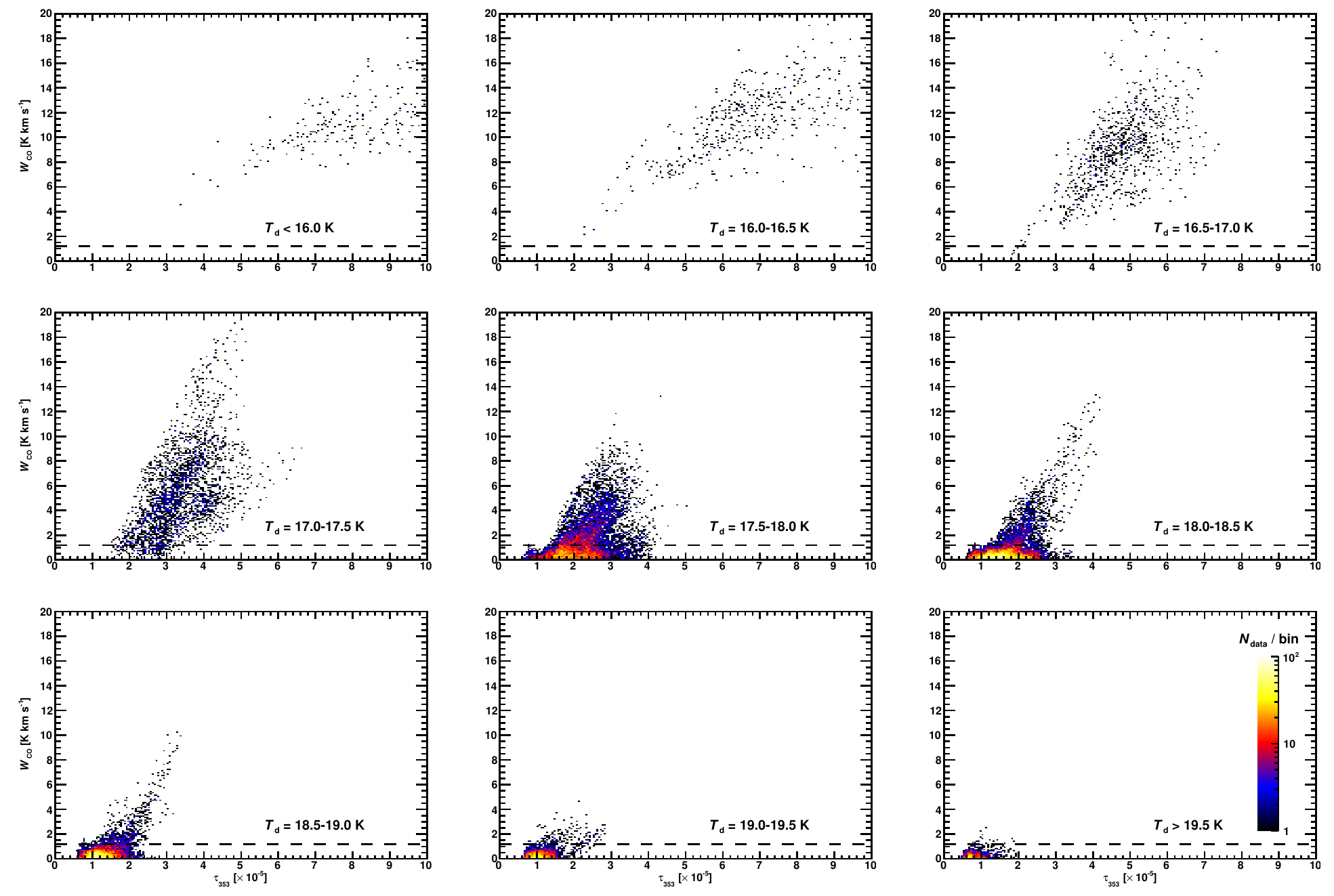}
  \end{center}
 \caption{Correlation plots (in density) between $\taud$ and $\WCO$ sorted by several $\Td$ intervals. The horizontal dashed lines indicate 3 $\sigma$ confidence level in $\WCO$.}
\label{fig:fig18}  
\end{figure*}

We present correlations of $\WHI$ and $\WCO$ with $\taud$ sorted by $\Td$ in Figures~\ref{fig:fig17} and \ref{fig:fig18}, respectively. 

\clearpage

\section{Derivation of Reference points in the $\NH$ Model} \label{sec:RefPoints}

The reference points in the $\NH$ model, $\taudref$ and $\NHref$ are determined in the same manner in \citet{Okamoto+17} for a study of the Perseus cloud.
Figure~\ref{fig:fig16}(a) shows that a scatter plot between $\Td$ and $\langle S_{i} \rangle$ (dispersion of each $\Td$ interval in the $\taud$--$\WHI$ plot).
The $\langle S_{i} \rangle$ is defined as a mean of the variance, which is derived from the areas of the right-angled triangles formed by each data point and regression line obtained by the fit to the data points in each $\Td$ interval
(cf., Figure 18 in \citealt{Okamoto+17}). 
As seen in the panel (a), the dispersion in $\Td$ tends to become small with increasing $\Td$.
The panel (b) represents a correlation between the averaged $\tauHI$ ($\langle \tauHI \rangle$) and $\langle S_{i} \rangle$ in each $\Td$ for the MBM~53--55 region.
We found a good positive correlation.
By using this correlation as a template, $\tauHI$ for the highest $\Td$ in the Chameleon region is derived to be 0.34.
Given a correlation, $\NHref = (1.15 \times 10^{8}) \times \XHI \times \taudref$, which is derived from the data points in the highest $\Td$ for |$b$| $>$ 15$^{\circ}$ in the all-sky data (F15), the model curve in Equation~(\ref{eq:WHIModelCurve}) with $\tauHI =$~0.34 gives $\taudref = 7.8\times10^{-7}$ and $\NHref = 1.6 \times 10^{20}$ cm$^{-2}$, thorough the fit to the data points only for the highest $\Td$ in the $\taud$--$\WHI$ relationship.
Table~\ref{tab:DispersionTauHIMBMPerseus} summarizes calculated dispersions in the $\taud$--$\WHI$ relationship for the Chamaeleon and MBM~53--55 regions.

\begin{figure*}[h]
 \begin{center}
  \includegraphics[width=120mm]{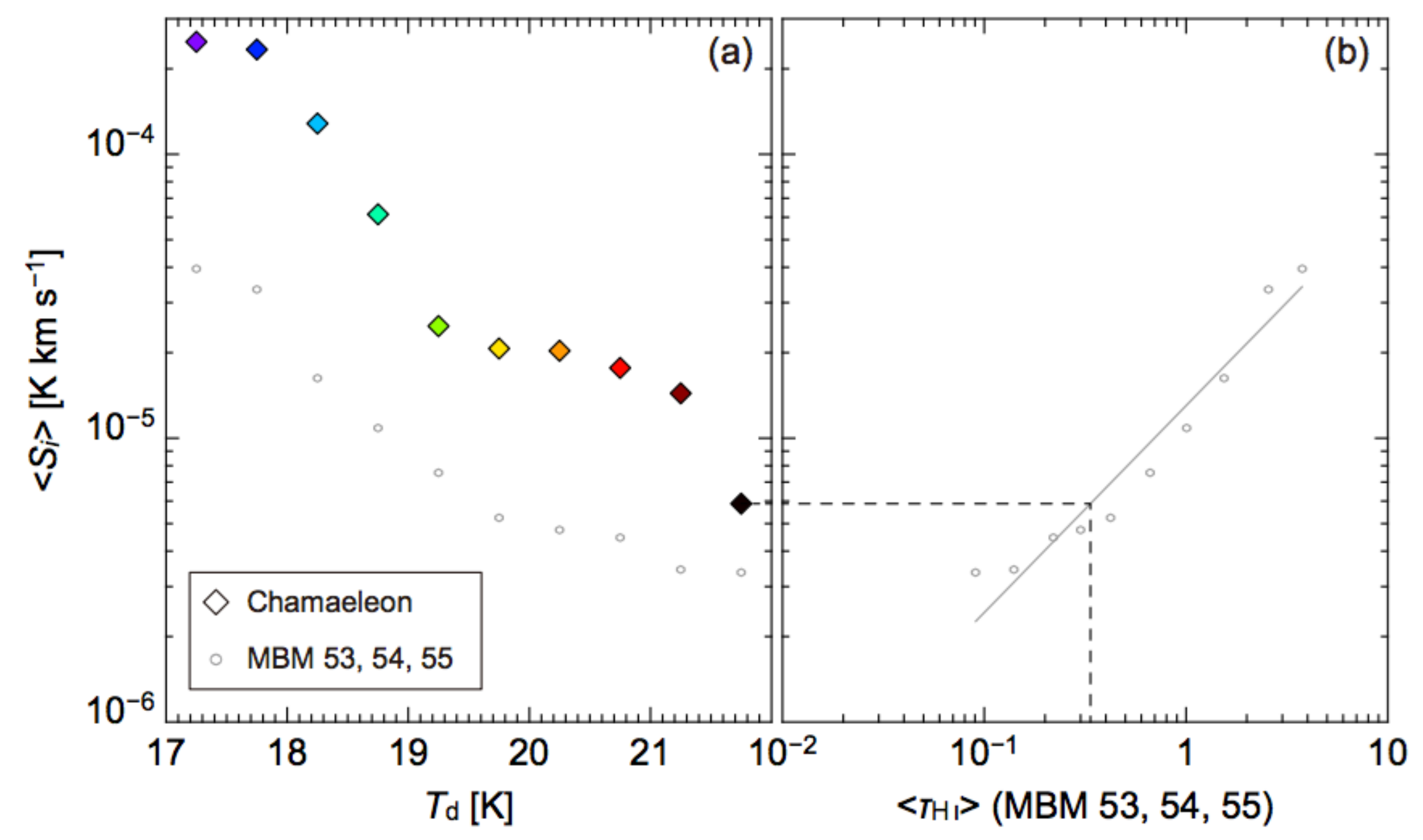}
  \end{center}
 \caption{(a) Correlations between $\Td$ and $\langle S_{i} \rangle$ (dispersion in the $\taud$--$\WHI$ relationship) for the MBM~53--55 and the Chamaeleon regions. For the Chamaeleon region, the data with different dust temperatures are shown by different colors adopted in Figure~\ref{fig:fig14}(c). $\langle S_{i} \rangle$ is defined as the mean area of the right-angled triangles formed by the data points and the regression line for each $\Td$ intervals \citep{Okamoto+17}. (b) The “template” relationship between $\langle \tauHI \rangle$ and $\langle S_{i} \rangle$ for the MBM~53--55 region. The solid line indicates the result of a linear regression. $\langle \tauHI \rangle$ for the highest-$\Td$ points in the Chamaeleon region can be estimated from $\langle S_{i} \rangle$ by using this “template”. The result gives $\langle \tauHI \rangle =$ 0.34.}
\label{fig:fig19}  
\end{figure*}

\begin{table*}[h]
\caption{Dispersion of several $\Td$ intervals in the $\taud$--$\WHI$ relationship for the Chamaeleon and MBM~53--55 regions.}
\label{tab:DispersionTauHIMBMPerseus}
 \begin{center}
\begin{tabular}{cccc} \hline\hline
    \makebox[6em][c]{} &  \makebox[9em][c]{Chamaeleon}     &   \multicolumn{2}{c}{MBM 53--55}  \\  \cline{3-4}
    $\Td$ & $\langle S_{i} \rangle$      & \makebox[5em][c]{$\langle S_{i} \rangle$}   & $\langle \tauHI \rangle$  \\ \
   (K)     & (10$^{-5}$ K km s$^{-1}$) &                                       & (10$^{-5}$ K km s$^{-1}$) \\ \
   (a)      & (b)                                     & (c)                                  & (d)                                     \\ \hline
   21.5 <  & 0.09                               & 0.59   & 0.34                                                                     \\
   21.0--21.5 & 0.14                               & 1.44 & 0.34                                                                 \\ 
   20.5--21.0 & 0.22                               & 1.77  & 0.45                                                                \\
   20.0--20.5 & 0.30                               & 2.03  & 0.48                                                                \\
   19.5--20.0 & 0.42                               & 2.07 & 0.52                                                                    \\
   19.0--19.5 & 0.66                               & 2.48 & 0.76                                                                     \\
   18.5--19.0 & 1.01                               & 6.14 & 1.09                                                                     \\
   18.0--18.5 & 1.54                               & 12.8  & 1.63                                                                   \\
   17.5--18.0 & 2.56                               &  23.4  & 3.34                                                                 \\
   < 17.5       & 3.76                               & 24.9  & 3.95                                                                    \\ \hline
\multicolumn{4}{l}{\scriptsize{(a) $\Td$ range}}\\
\multicolumn{4}{l}{\scriptsize{(b) Dispersions of each $\Td$ for the (b) Chamaeleon region.}}\\
\multicolumn{4}{l}{\scriptsize{(c) The same as column (b) but for the MBM~53--55 region.}}\\
\multicolumn{4}{l}{\scriptsize{(d) $\HI$ optical depth $\langle \tauHI \rangle$ for the MBM~53--55 region in each $\Td$ (Table 3 in \citealt{Okamoto+17}).}}\\
\end{tabular}
\end{center}
\end{table*}

\clearpage

\section{$\XCO$ Distribution} \label{sec:XcoAppendix}

\begin{figure*}[h]
 \begin{center}
  \includegraphics[width=180mm]{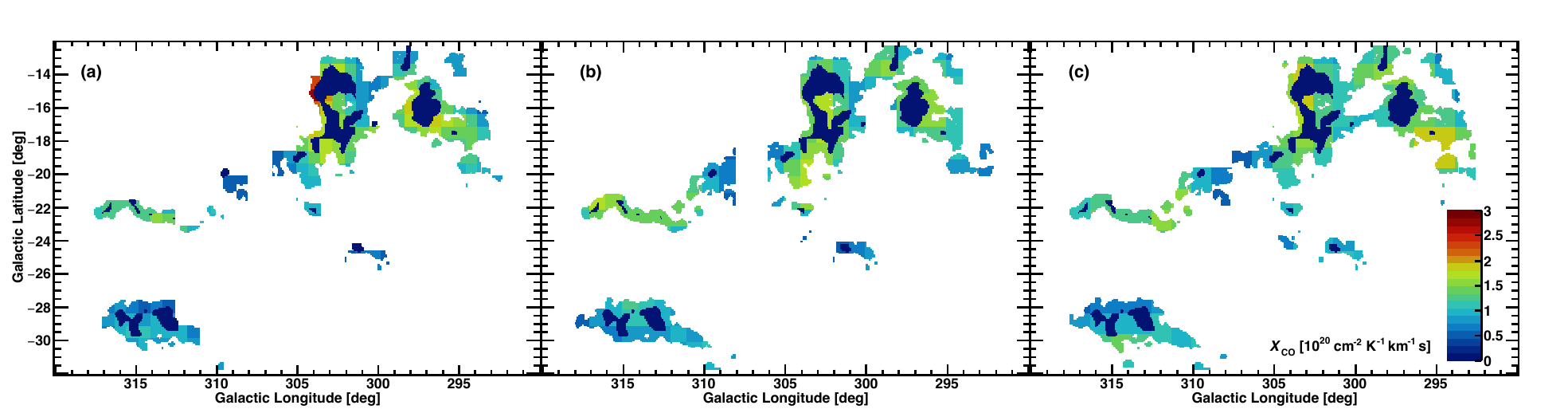}
  \end{center}
 \caption{$\XCO$ distribution derived from the correlation between $\WCO$ and $\NH$ (see Section~\ref{sec: Xco}): (a) 
1$\fdg$5 $\times$ 1$\fdg$5-based map smoothed with a two-dimensional Gaussian function with a kernel size of three pixels and $\sigma=1^{\circ}$; (b) 2$^{\circ}$$\times$2$^{\circ}$-based map with a Gaussian function of $\sigma=1.3^{\circ}$; (c) 2$\fdg$5 $\times$ 2$\fdg$5-based map with a Gaussian function of $\sigma=1.7^{\circ}$. The grid sizes are (a) 0$\fdg$75 $\times$ 0$\fdg$75, (b) 1$\fdg$0 $\times$ 1$\fdg$0 and (c) 1$\fdg$25 $\times$ 1$\fdg$25.}
\label{fig:fig20}  
\end{figure*}

We present the obtained $\XCO$ maps with the different grid sizes in Figure~\ref{fig:fig20}.

\end{document}